\documentclass[fleqn,usenatbib]{mnras}

\usepackage{newtxtext,newtxmath}

\usepackage[T1]{fontenc}
\usepackage{gensymb}
\usepackage{csvsimple}
\usepackage{longtable}
\usepackage{lscape}
\usepackage{multicol}
\usepackage{multirow}
\usepackage{subcaption}
\usepackage{supertabular}
\usepackage{rotating}
\usepackage{placeins}
\usepackage{anyfontsize}
\usepackage{booktabs}
\usepackage{enumerate}

\DeclareRobustCommand{\VAN}[3]{#2}
\let\VANthebibliography\thebibliography
\def\thebibliography{\DeclareRobustCommand{\VAN}[3]{##3}\VANthebibliography}

\usepackage{graphicx}	
\usepackage{amsmath}	

\newcommand\hi{\mbox{H\,{\sc i}}\,}
\newcommand\hix{\mbox{H\,{\sc i}}}
\newcommand\hiit{\mbox{H\,{\textsl{\textsc{i}}\,}}}

\title[\hiit absorption in star-forming galaxies. I. ]{A search for \hi absorption in distant star-forming galaxies with ASKAP-FLASH - I. Selection and analysis of the radio sample}

\author[S. L. Eden et al.]{
Sophie L. Eden,$^{1}$\thanks{E-mail: s.eden-2017@hull.ac.uk}
Elaine M. Sadler,$^{2,3,4}$
Kevin A. Pimbblet,$^{1}$
Elizabeth K. Mahony,$^{5}$
Hyein Yoon,$^{6,7,2}$
\\
$^{1}$E. A. Milne Centre for Astrophysics, University of Hull, Cottingham Road, Kingston-upon-Hull, HU6 7RX, UK \\
$^{2}$Sydney Institute for Astronomy, School of Physics A28, University of Sydney, Sydney, NSW 2006, Australia \\
$^{3}$ATNF, CSIRO Space and Astronomy, PO Box 76, Epping, NSW 1710, Australia\\
$^{4}$ARC Centre of Excellence for All Sky Astrophysics in 3 Dimensions (ASTRO 3D) \\
$^{5}$ATNF, CSIRO Space and Astronomy, PO Box 76, Epping, NSW 1710, Australia \\
$^{6}$Institute for Data Innovation in Science, Seoul National University, 1 Gwanak-ro, Gwanak-gu, Seoul 08826, Korea \\
$^{7}$Astronomy Program, Department of Physics and Astronomy, Seoul National University, 1 Gwanak-ro, Gwanak-gu, Seoul 08826, Korea \\
}

\date{Accepted XXX. Received YYY; in original form ZZZ}

\pubyear{2024}

\begin{document}
\label{firstpage}
\pagerange{\pageref{firstpage}--\pageref{lastpage}}
\maketitle

\begin{abstract}

\noindent
We present and discuss two catalogues of UV-selected (NUV $< 22.8$\,mag) galaxies that lie within a 200 deg$^2$ area of sky covered by the ASKAP FLASH survey and have an impact parameter of less than 20\,arcsec to a FLASH radio continuum source. These catalogues are designed to enable a future search for 21\,cm \hi absorption in and around star-forming galaxies at redshift $0.4<z<1$. We outline the production of this UV-bright dataset, which has optical spectroscopy from the WiggleZ and SDSS surveys and a median redshift of $\sim0.6$. Analysis of the optical spectra, using multiple diagnostic diagrams, shows that galaxies with an impact parameter of less than 5\, arcsec are likely to be physically associated with the radio source and are five times more likely to be an AGN than objects without a radio match. Conversely, objects with impact factors between 5 and 20\, arcsec are largely (>\,80\,percent) star-forming and resemble the overall WiggleZ population. The ($g - i$) colour evolution with redshift is consistent with a history of active star-formation, but the radio-associated objects are typically redder and have colours similar to high-excitation radio galaxies. The redshift distribution of the two catalogues matches the overall distribution for WiggleZ galaxies, despite their otherwise rare radio properties. These catalogues can be expanded in future as new radio data become available, and a forthcoming paper will present the \hi absorption results. 

\end{abstract}




\section{Introduction}\label{intro}

There is a well-observed decline in the global star-formation rate by almost a factor of $\sim$10 between `cosmic noon' - where global SFR hit its peak at \textit{z} $\sim{2.0}$ - and present day, referred to as `cosmic downsizing' \citep{Cowie1996,Madau1996,Madau2014,Cochrane2023}. The reasons for this phenomenon are not well understood, but it may be connected to the evolution of the cosmic neutral hydrogen abundance over time.

Neutral hydrogen is a primary constituent of the reservoirs of cool matter surrounding galaxies \citep{Emonts2007,Kaufmann2006}, which fuel stellar formation \citep{Sancisi2008,Stroe2015} and active galactic nuclei (AGN) evolution \citep{Gereb2014,Stroe2015}. If \hi can be probed via 21\,cm absorption detected in galaxies across a large range of redshifts, it offers the opportunity to study the rates of \hi detection over cosmic time. If we can specifically limit our sample to star-forming galaxies, we can also investigate the role of \hi in star formation, or alternatively the impact of star formation on \hi.

Though we expect that \hi reservoirs should be detected around star-forming galaxies, due to the UV-luminosity produced by massive stars, there is also a possibility that even after running spectral stacking experiments, detections will not occur at all, or only for galaxies where the production of ionising UV radiation - and likely the star formation rate - is low enough \citep{Curran2019}. This would offer the opportunity to analyse the feedback between \hi and star formation specifically, while excluding contamination from AGN emission.

\hi is prevalent throughout the Universe \citep{Wolfe2005,Cortese2017,Hu2020}, and produces an emission line at 21\,cm (1.42\,GHz). This hyperfine transition is very faint, limiting associated detection of the 21\,cm emission to galaxies with redshift \textit{z}\,<\,0.4 \citep{Fernandez2016}, except in exceptional cases such as the \citet{Chakraborty2023} observation at \textit{z}\,$\sim{1.3}$ which was enabled by gravitational lensing of the source.

The study of \hi emission using the 21\,cm line is generally restricted to the local Universe and low redshift regions. Stacked \hi emission signals may be detected at higher redshift, but this is difficult, requiring many \hi targets to increase signal-to-noise, and long integration times \citep{Rhee2018,Chowdhury2021}. Damped Ly\,$\upalpha$ absorption lines can be effective tracers of distant \hi \citep[e.g.][]{Wolfe2005,noterdaeme12,
zafar13}, but it is only detectable from the ground for \textit{z}\,>\,1.7. At lower redshift, this line lies in the UV and can only be observed from space. Because of this, the current samples of lower-redshift quasar DLAs are small and may suffer from a range of selection effects \citep{neeleman16,rao17}. 

Thus, for the redshift range 0.4\,<\,\textit{z}\,<\,1.7, there is currently an incomplete understanding of the evolution of neutral hydrogen through time. Surveys for 21\,cm \hi absorption against background radio sources provide an alternative probe \citep{Grasha2020,Gupta2021}, and a new generation of wide-band SKA precursor telescopes on radio-quiet sites has opened a new parameter space for such studies \citep{Allison2022}. 

As star formation and AGN activity are expected to be linked to the quantity of neutral gas in and around a galaxy, searches for 21\,cm \hi absorption on sightlines near galaxies with a high SFR or hosting AGN can provide new insights into the properties of \hi in distant galaxies. These galaxies would be UV-bright \citep{Smith1989,Tadhunter2002}, and so can be identified from large-area imaging surveys like GALEX \citep{Martin2005}.

We aim to produce a catalogue of UV-bright galaxies suitable for a large-area \hi search with the ASKAP radio telescope \citep{Hotan2021}. We use objects from the WiggleZ survey \citep{Drinkwater2010,Drinkwater2018}, due to its selection of objects with NUV\,<\,22.8\,mag, 90\,percent of which lie in the redshift range 0.2\,<\,\textit{z}\,<\,1.0, and 90\,percent of which are star-forming, across a total area of 1000\,deg$^{2}$ of sky. A substantial fraction of the WiggleZ survey area also has existing \hi absorption data from the ASKAP-FLASH survey \citep{Allison2022}, covering the redshift range 0.4\,<\,\textit{z}\,<\,1.0.

In this paper we describe the process of producing catalogues of optical objects in the redshift range 0.4\,<\,\textit{z}\,<\,1.0 which either host - or alternatively are in the foreground of - radio sources detected by ASKAP-FLASH. As part of this work, we extend the initial catalogue of optical objects by producing an additional field of `WiggleZ-type' objects which match the WiggleZ selection criteria, and show that the resulting objects match the properties of WiggleZ objects well.

We investigate the properties of our two radio cross-matched subsamples of objects, and their similarity to the WiggleZ survey as a whole by investigating their optical emission line properties, redshift distributions, and (\textit{g - i}) colour evolutions. This allows us to evaluate the suitability of our data for this future \hi absorption work, which would be some of the first done at these redshifts. The catalogue presented here will later be used to search FLASH spectra for 21\,cm absorption signals, both via associated detections and through absorption-line stacking experiments to statistically test for the presence of \hi even when column densities would otherwise be too low for a direct detection to be made. We are also interested in the possibility of investigating ionisation of neutral gas due to the ionising emission of massive stars in star-forming galaxies, which could impact potential \hi existing around these galaxies.

Section \ref{surveys} discusses the surveys of interest for this work: WiggleZ optical data, FLASH radio data, and outlines the production of our extended sample data in accordance with WiggleZ selection criteria \citep{Drinkwater2010}. In Section \ref{cross-matching} we cover our optical-radio cross-matching approach, and the two separate populations of interest we produce via this process. We analyse the optical spectra of our overall catalogue of objects, and our two cross-matched samples, by producing BPT and WHAN spectral line diagnostic diagrams in Section \ref{bpt_whan}. In Section \ref{discussion} we do further studies of our objects, looking at colour evolution over redshift, redshift distribution, and MEx data, while comparing against existing literature, before concluding our findings in Section \ref{conclusion}. Sample entries of the data produced in this work are presented in the Appendix, as well as example spectra. Throughout this work, we assume a concordance cosmology model where $\upOmega_{\upLambda}$\,=\,0.7, $\upOmega_{\mathrm{m}}$\,=\,0.3, and H$_{0}$\,=\,70\,km\,s$^{-1}$\,Mpc$^{-1}$.


\section{Survey Area}\label{surveys}

\begin{figure*}
    \centering
    \includegraphics[width=\linewidth]{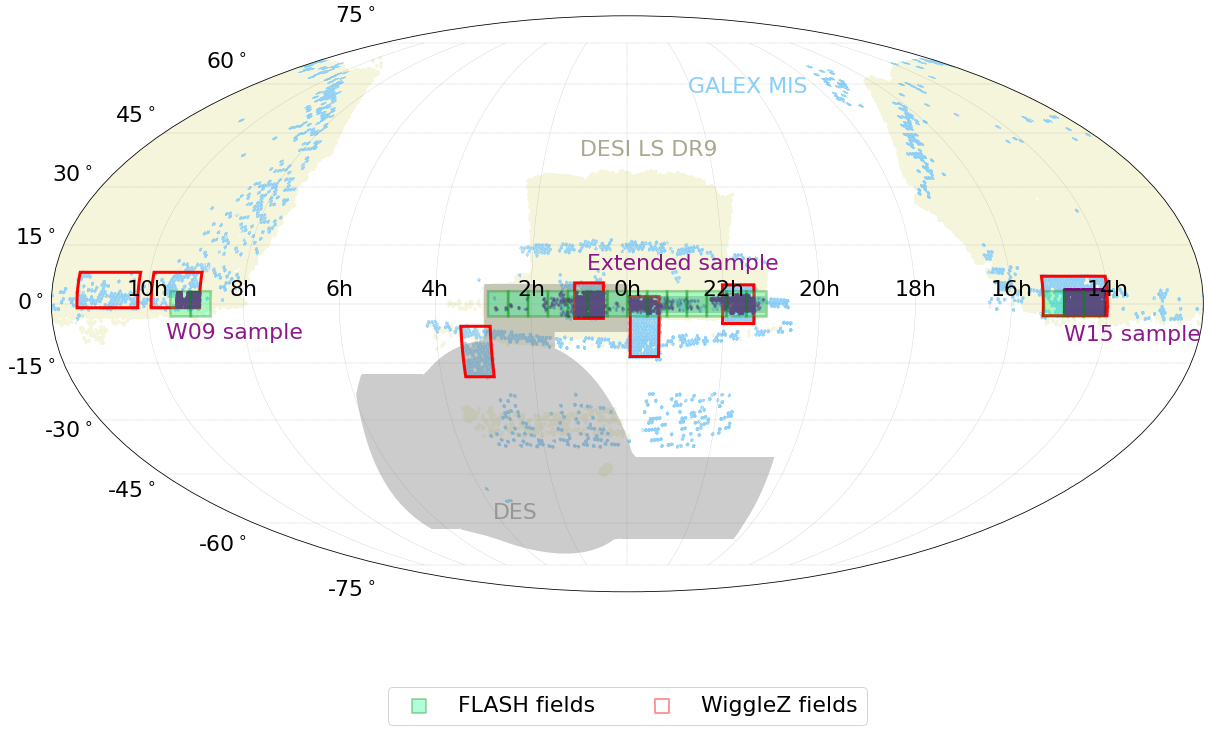}
    \caption{Footprint plot of DES (grey), DESI Legacy Survey DR9 (yellow), and GALEX MIS (light blue) data. WiggleZ (red outline) and FLASH fields (green) overlaid. W09, W15, and resulting extended sample fields are shown in purple. It can be seen that the W09 and W15 samples used in this paper do not occupy the entire WiggleZ 9h and 15h fields. The total coverage of the sky in this work is 215\,deg$^{2}$. See Table \ref{tab:areas_table} for detail on field limits.}
    \label{fig:footprint}
\end{figure*}


\subsection{The WiggleZ spectroscopic survey}\label{wigglez}

\subsubsection{Background}
The WiggleZ survey \citep{Drinkwater2010,Drinkwater2018} was originally designed to enable a cosmological study of Baryon Acoustic Oscillations (BAO) by providing 
a catalogue of over 240,000 emission-line redshifts for distant, star-forming galaxies. The main redshift range covered is 0.2\,<\,\textit{z}\,<\,1.0, so WiggleZ reaches to significantly higher redshifts than other large-area spectroscopic surveys like SDSS and GAMA. 

The galaxies included in the WiggleZ survey were selected to have a UV luminosity typical of galaxies with a high star-formation rate \citep{Drinkwater2018}. GALEX Medium Imaging Survey (MIS) UV photometry was used to select candidate objects with NUV\,<\,22.8\,mag. The GALEX objects were then cross-matched with the SDSS and RCS2 optical catalogues to provide more accurate 
positions for spectroscopic observation \citep{Drinkwater2010}.

\subsubsection{Sky area}
The $\sim$1000\,deg$^2$ area of sky covered by WiggleZ is split into seven separate fields (see table\,1 of \cite{Drinkwater2018} for details). They include the three equatorial GAMA fields \citep{Driver2011,Liske2015} for which SDSS photometry is available, along with four additional fields with optical photometry from the Canada–France–Hawaii Telescope (CFHT) Second Red-sequence Cluster Survey \citep[RCS2;][]{Gilbank11}. 

The main focus of this paper is on the WiggleZ SDSS regions, which overlap with radio data from the ASKAP FLASH Pilot Survey fields (Yoon et al., in prep.). We began with the overlapping part of WiggleZ 15h field, before extending out into the additional area covered by the WiggleZ 9h field. Our two WiggleZ regions of interest are displayed on Fig.\,\ref{fig:footprint} as solid purple regions occupying smaller areas within their respective WiggleZ fields outlined in red. As the overlap regions of these fields used in our work are smaller than the original WiggleZ fields, they will be referred to as W09 and W15 in this paper.

As discussed below in section \ref{wigglez-like}, we extend our WiggleZ-FLASH sample to cover additional UV-bright galaxies in equatorial fields covered by the WiggleZ 22h, 0h and 1h and areas in, using the WiggleZ survey design and selection criteria outlined in \citet{Drinkwater2010}. 

\begin{table}
\centering
\begin{tabular}{l|r|r|r}
    \hline \hline
    Source Class & N$_{\mathrm{9h}}$ & N$_{\mathrm{15h}}$ & N$_{\upSigma}$\\
    \hline    
    WIG\_SDSS & 5813 & 18388 & 24201\\
    WIG\_SDSS\_EXT & 1050 & 2687 & 3737\\
    SPARE\_RADIO & 184 & 374 & 558\\
    SPARE\_GALEX & 14 & 12 & 26\\
    \hline
    Total & 7061 & 21461 & 28522\\
    \hline
\end{tabular}
\caption{The class designations of WiggleZ objects within the subsections of the WiggleZ 9h and 15h fields used in this paper, and quantity of each class. It can be seen that for these portions of the W09 and W15 WiggleZ fields, the vast majority of objects are classified as WIG\_SDSS, which are SDSS galaxies that meet all final criteria of the WiggleZ survey. WIG\_SDSS\_EXT are SDSS objects that were observed early in the WiggleZ run, but did not satisfy final criteria. SPARE\_RADIO and SPARE\_GALEX are examples of object classes that were studied if spare fibres were available during an observing run.}
\label{tab:wigglez_source_types}
\end{table}

\subsubsection{WiggleZ spectra}

The WiggleZ objects have designated classes, examples of which can be seen in Table \ref{tab:wigglez_source_types}. This table only includes classes for objects in the W09 and W15 fields. The main set of objects that we are interested in are the WIG\_SDSS objects which comprise the majority of the WiggleZ 15h field, and roughly a quarter of WiggleZ 9h field, becoming the W15 and W09 fields respectively.

The SPARE\_RADIO class consists of objects that \citet{Ching2017} identified as part of their LARGESS survey. Their optical spectra were observed by WiggleZ in their effort to observe for companion projects if they were unable to use all fibres on high priority WiggleZ targets. These are radio galaxies with an optical component, however, these galaxies are not necessarily UV-bright, and have a different selection criteria from WiggleZ objects. For this reason, objects with the SPARE\_RADIO class were not considered for this work.

The optical spectra of WiggleZ galaxies are publicly available, and some examples are shown in Fig. \ref{fig:sf_spectrum}, \ref{fig:seyfert_spectrum} and \ref{fig:comp_spectrum}. 
Each spectrum observed in the WiggleZ survey has an associated `QOP' value assigned, which indicates the quality of the listed spectroscopic redshift. This QOP value relies on specific lines seen in spectra, and the fitting of templates of typical types of galaxies, in order to judge whether the redshift is accurate. In this paper, we only consider WiggleZ galaxies with QOP values between 3 and 5, where 3 indicates a reasonably reliable redshift, and 5 an ideal spectrum that could itself be used as a template. This ensures we only select galaxies with a high likelihood of possessing a reliable spectroscopic redshift. 

Spectroscopically, the WiggleZ galaxies are almost entirely star-forming systems rather than AGN (from table 8 of \cite{Drinkwater2018}, fewer than 10\,percent show a spectroscopic AGN signature). It therefore seems likely that most of these galaxies possess a reservoir of neutral hydrogen and are likely to contain \hi disks \citep{deBlok2003,Bigiel2010,Yildiz2017}. 

\subsubsection{Photometry and additional data}
The WiggleZ catalogue also includes GALEX MIS UV photometry and optical photometry in the \textit{g}, \textit{r}, and \textit{i} bands. These are especially useful for analysis complementary to the diagnostic analyses their spectra allow, and will be explored later in Section \ref{bpt_whan} and \ref{discussion}. The catalogue also provides stellar mass estimates for most WiggleZ galaxies. To produce these data, WiggleZ optical and UV photometry were used with the KG04 spectral fitting code \citep{Glazebrook2004}, PEGASE.2 stellar models \citep{Fioc1997,Fioc1999}, and a \citet{Baldry2003} initial mass function to produce mass estimates for 82\,percent of the galaxies in the WiggleZ catalogue - see \citet{Drinkwater2018} for further detail.

These estimates will be useful for further detailed analysis of objects where available in our extended catalogue that are WiggleZ galaxies - i.e. that either originated in the WiggleZ catalogue and are found in W09 and W15, or that lie in the extended field and contain WiggleZ spectroscopic redshift as found in DR9 - and will be useful for analysis in future work. However in this work, we do use stellar masses for an MEx diagnostic diagram in Section \ref{mex}.


\subsection{FLASH}\label{flash}
\subsubsection{Background and Pilot Survey area} 

The FLASH survey \citep{Allison2022} is using the ASKAP telescope to search for 21\,cm absorption in the redshift range 0.42\,<\,\textit{z}\,<\,1.0. While the survey will eventually cover the whole southern sky, the focus of this paper is the $\sim$3200\,$\mathrm{deg^{2}}$ area already observed in the FLASH Pilot Survey (Yoon et al. \textit{in prep.}), as shown in Fig.\,\ref{fig:footprint}. 

ASKAP data products are produced via the automated \verb/ASKAPsoft/ pipeline (\citeauthor{Whiting2020} \citeyear{Whiting2020}; Yoon et al. \textit{in prep.}), which allows extraction of useful data from observations, without storing raw data which are prohibitively large. Source finding algorithms are used to produce both component catalogues and island catalogues (by calculating the likelihoods of association between components in the 3D sky, and flagging whatever source appears to be at the group's centre), using \verb/Selavy/ \citep{Whiting2012b}. 

Spectra are then extracted at the positions of several hundred bright radio sources in each field, and the \verb/FLASHfinder/ line-finding tool \citep{Allison2012} is used to search for \hi absorption lines. 

FLASH has already demonstrated its ability to detect `intervening absorbers' - \hi objects illuminated by continuum radio sources behind them by line of sight to the observer \citep{Allison2020,Sadler2020} - which is what this work seeks to do with the specific population of WiggleZ galaxies.

\subsubsection{The FLASH continuum data}

The radio continuum catalogues produced by the FLASH survey have a central frequency of 856\,MHz and 
are both deeper (rms noise $\sim0.1$\,mJy\,beam$^{-1}$), and have a higher spatial resolution ($12-15$\,arcsec beam) than the large-area NVSS, SUMSS, and RACS surveys, making them ideal for identifying radio continuum sources within and near WiggleZ galaxies. 

The FLASH pathfinder fields cover a large portion of the celestial equator, and so overlap the W09 and W15 WiggleZ fields, as well as the extended sample (see Section \ref{wigglez-like} for details), allowing positional cross-matching between the WiggleZ-type UV-bright objects, and radio continuum sources. Table \ref{tab:flash_fields_table} contains a list of all FLASH fields relevant to this paper, and these fields are displayed on Fig.\,\ref{fig:footprint} to show their sky coverage.

\begin{table}
\centering
\begin{tabular}{l|l}
    \hline \hline
    Field & FLASH pilot survey fields\\
    \hline    
    W09 & 546, 547 \\
    W15 & 559, 560, 561 \\
    Extended Sample & 525, 526, 527, 528, 529, 530, 531, 576, 577, 578,\\ 
     & 579, 580, 581, 582\\
    \hline
\end{tabular}
\caption{List of FLASH fields providing coverage for each field of objects satisfying selection criteria in this paper and displayed on Fig.\,\ref{fig:footprint}.}
\label{tab:flash_fields_table}
\end{table}

In order to mitigate the impact of noise at the edges of FLASH fields where coverage is worst and noise is highest, FLASH fields are tiled, with the edges of the fields overlapping where coverage is worst and noise is highest. For this reason, before FLASH data was cross-matched the catalogues of FLASH fields used in this work (see Fig.\,\ref{tab:flash_fields_table}) were trimmed down to leave only the central 6.4$^{\circ}$ by 6.4$^{\circ}$ region at the centre of each field, where noise is lowest, and essentially remains uniform. This also means there should be no duplication of sources between one field and another.


\begin{figure*}
    \centering
    \includegraphics[width=\linewidth]{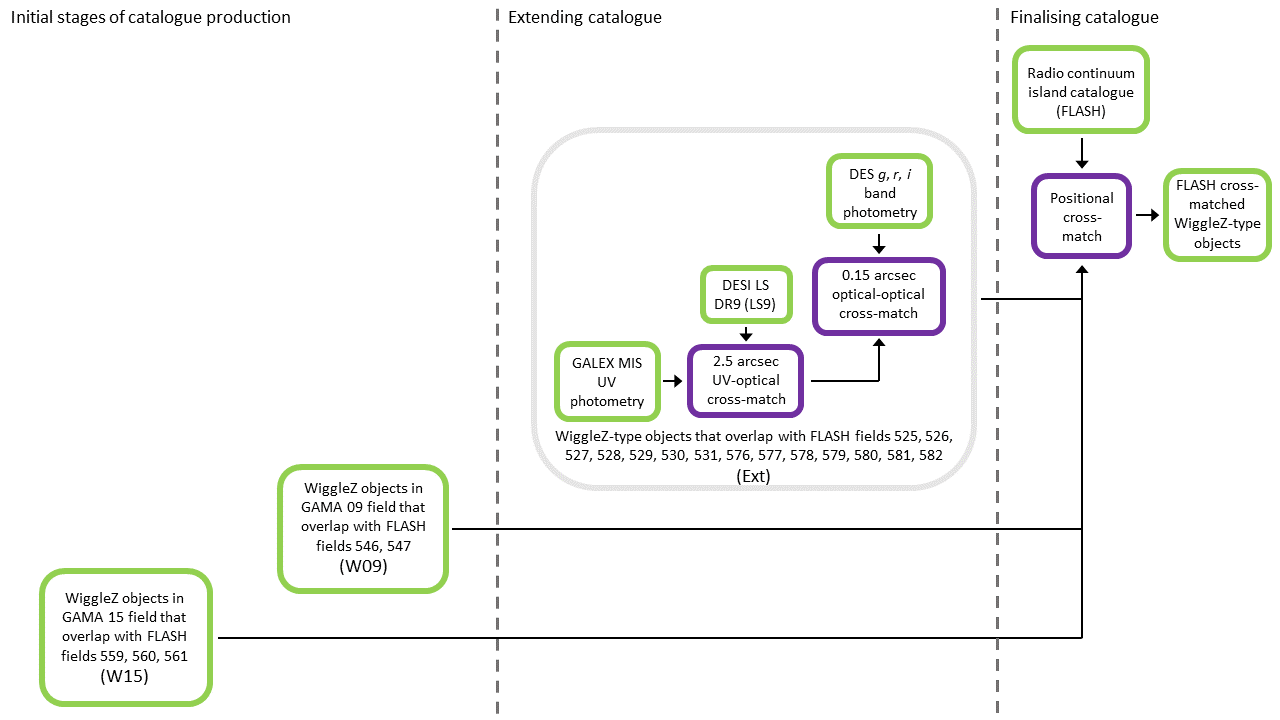}
    \caption{Visualisation of catalogue production, split into stages of initial approach by vertical grey dashed lines: extending the catalogue, and lastly obtaining the finalised catalogue. The specific steps of production of the Ext field objects is outlined within the pale grey bubble.}
    \label{fig:sample_flowchart}
\end{figure*}

\subsection{An extended WiggleZ-like catalogue}\label{wigglez-like}

Alongside the objects in the W09 and W15 fields, we desired a larger catalogue of WiggleZ-like objects, with one of the goals of our future \hi work being spectral stacking experiments, where larger datasets help increase the likelihood of detection of 21\,cm signals by driving down noise in the stacked spectrum. In order to extend this work to a larger sample, we added additional catalogue data within the area covered by the FLASH Pilot Survey. A visualisation of the production of this extended data sample and its relation to the W09 and W15 fields is included in Fig.\,\ref{fig:sample_flowchart}.

The additional catalogue data were produced using the Dark Energy Spectroscopic Instrument (DESI) Legacy Survey DR9 \citep{Dey2019}, Dark Energy Survey (DES) DR2 \citep{Abbott2021}, and GALEX MIS UV survey \citep{Martin2005}. DESI LS DR9 (hereon LS9) is an optical survey containing spectroscopic redshift data, allowing a sample of objects with redshifts between 0.4\,<\,\textit{z}\,<\,1.0 to be produced. As LS9 does not contain all three \textit{gri} photometric bands, DES data was folded in to provide additional photometry for direct comparison with WiggleZ objects. 

We used GALEX MIS photometry to put in place the UV cutoffs required, and we excluded objects that did not show an FUV dropout (see table\,4 of  \citet{Drinkwater2010} for a full breakdown of the WiggleZ selection criteria). As GALEX MIS is limited to 1000\,deg$^{2}$ of the sky, the possible area that any resulting WiggleZ-like catalogue can cover that is not already covered by WiggleZ, is limited.

LS9 lists the surveys that acted as sources for the spectroscopic redshifts of each object included in it. These redshifts are obtained from external surveys, and 85\,percent of LS9 objects in the extended field possess a spectroscopic redshift from WiggleZ with the remained being from BOSS/eBOSS. 

This is due to the external field covering parts of the 0h, 1h, and 22h WiggleZ fields. These galaxies are WiggleZ galaxies that used RCS2 photometry instead of SDSS photometry, and so were excluded by us when we first limited our objects to WIG\_SDSS class objects. The only data for these objects retained from the WiggleZ survey in this work is their spectroscopic redshift (as found in LS9), and their WiggleZ spectra, and stellar mass values, which are used in later analyses.

To produce a catalogue from these three disparate datasets, we carried out positional cross-matching using \verb\Aladin Sky Atlas\ \citep{Bonnarel2000}, and followed the WiggleZ selection criteria from \citet{Drinkwater2010}. First, we cross-matched GALEX MIS and LS9 data out to a radius of 2.5\,arcsec in order to produce a set of UV-optical objects with spectroscopic redshift. Next, we matched the LS9 optical component to DES optical data, out to a radius of 0.15\,arcsec, to include the \textit{gri} photometry we require. Once a catalogue of UV-optical objects was produced, the photometric cuts of the WiggleZ selection criteria were imposed.

As can be seen in Fig.\,\ref{fig:footprint}, the extended sample data was limited to areas of the sky where LS9, DES, and GALEX coverage overlapped. Thus, it was restricted to an equatorial strip from 3h to 21h, wrapping around 0h. Table \ref{tab:areas_table} outlines the extent and total area covered for the three fields analysed in this work. 

 \begin{table}
\centering
\begin{tabular}{l|r|r|r|r|r|r}
    \hline \hline
    Field & RA$_{\rm min}$ & RA$_{\rm max}$ & Dec$_{\rm min}$ & Dec$_{\rm max}$ & Area & N\textsubscript{$\upSigma$(field)}\\
     & (deg) & (deg) & (deg) & (deg) & (deg$^{2}$) \\
    \hline    
    W09 & 133.7 & 150.0 & $-0.9$ & $+3.0$ & 23.2 & 5813\\
    W15 & 211.0 & 223.5 & $-2.7$ & $+3.7$ & 69.5 & 18388\\
    Ext & 316.5 & 43.5 & $-3.1$ & $+3.1$ & 122.3 & 7462\\
    \hline
\end{tabular}
\caption{Position on sky, areas of field regions, and total objects fulfilling WiggleZ selection criteria contained within them, pre positional cross-matching. Note that the extended sample field wraps around 0h. The total sky area covered is 215\,deg$^2$. }
\label{tab:areas_table}
\end{table}

Additionally, redshift quality flags were necessary for all data in the extended sample. All objects with spectroscopic redshifts from the BOSS/eBOSS surveys use a `zwarning' category of 0 to flag good quality, reliable spectroscopic data. Further detail is given by \citet{SDSS}. As all BOSS/eBOSS objects are flagged as possessing reliable spectroscopic redshifts, our confidence in using these redshifts is high.


\subsection{UV-optical catalogue properties}\label{uv_properties}

\subsubsection{Redshift distribution}
As the UV-optical objects in this work are from WiggleZ data (W09 and W15 fields), and extra object data produced from a combination of other surveys with photometric selection criteria that matches the WiggleZ survey applied (extended field), it is expected that the behaviour of all three samples should be very similar. The structure of the WiggleZ survey prioritises not only UV-bright objects, but UV-bright objects at higher redshifts, resulting in a median of \textit{z}\,$\sim{0.6}$.

\begin{figure}
    \centering
    \includegraphics[width=\linewidth]{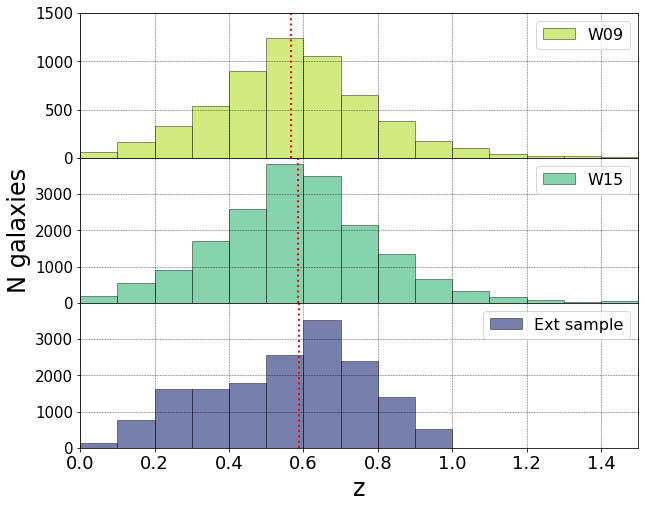}
    \caption{Redshift distribution of WiggleZ-type UV-optical cross-matched sources for the W09, W15, and extended sample fields with UV and optical photometric cuts enacted. Median redshift is marked with a red dotted line on each histogram. Despite the different redshift distribution of the extended sample, its median redshift is very close to that of the W09 and W15 fields, and that of the target median redshift of the WiggleZ survey (\textit{z}\,$\sim{0.6}$). Median redshifts for W09, W15, and extended sample are 0.566, 0.584, and 0.588 respectively. Note that the extended sample was limited to spectroscopic redshift <\,1.0, therefore its data cuts off above this point. Its WiggleZ object component has a median redshift of 0.558, while its component produced from all other spectroscopic surveys has a median redshift of 0.721.}
    \label{fig:redshift_hist}
\end{figure}

Fig.\,\ref{fig:redshift_hist} shows the redshift distribution of the W09, W15, and extended sample data once photometric cuts are in place. The W09 and W15 data show extremely similar distributions, with data peaking in the 0.4\,<\,\textit{z}\,<\,0.6 bin and a noticeable tail of a small number of objects at \textit{z}\,<\,1.0. On the other hand, the extended sample is slightly different, peaking in the 0.6\,<\,\textit{z}\,<\,0.8 bin, and possessing no data above \textit{z}\,=\,1.0.

The lack of data above \textit{z}\,=\,1.0 is due to the spectroscopic LS9 redshifts being limited to between 0\,<\,\textit{z}\,<\,1.0. However, the exact reason for the peak of the extended sample's redshift distribution being at a higher redshift is not known, though it may be due to the overall LS9 survey having a higher target median redshift than the overall WiggleZ survey. The median redshift for all three populations is \textit{z} $\sim{0.6}$ (dotted red line on Fig.\,\ref{fig:redshift_hist}), which as noted above, is the median of the full release WiggleZ survey.

Later in this work we show that the extended sample otherwise shows very similar behaviour to the WiggleZ objects, which makes the difference in redshift distribution interesting. However, given the closely matching median redshifts we see here, along with the objects of most interest to us being between 0.4\,<\,\textit{z}\,<\,1.0, we do not expect the difference in redshift distribution between samples to have a large impact on this work.


\section{Cross-matching with FLASH radio sources}\label{cross-matching}

\subsection{Cross-matching strategy}\label{cross-matching_strategy}

In order to distinguish between genuine matches and chance alignments on the sky (offset sources), the maximum distance between a FLASH radio source and its optical host galaxy must be found. Previous works such as \citet{Best2005b,Mauch2007,Ching2017} primarily work with the NVSS \citep{Condon1998} and FIRST \citep{Becker1995} radio surveys, producing catalogues of cross-matched radio sources and optical spectroscopic objects, which then allow analysis of the environments these radio objects are in, and how they evolve through time.

These prior works used radio component catalogues, which required the authors to associate separate components into extended morphology radio objects. In the case of FLASH, this is not required of the user, as island catalogues of associated components are also available. 

\begin{figure}
    \centering
    \includegraphics[width=\linewidth]{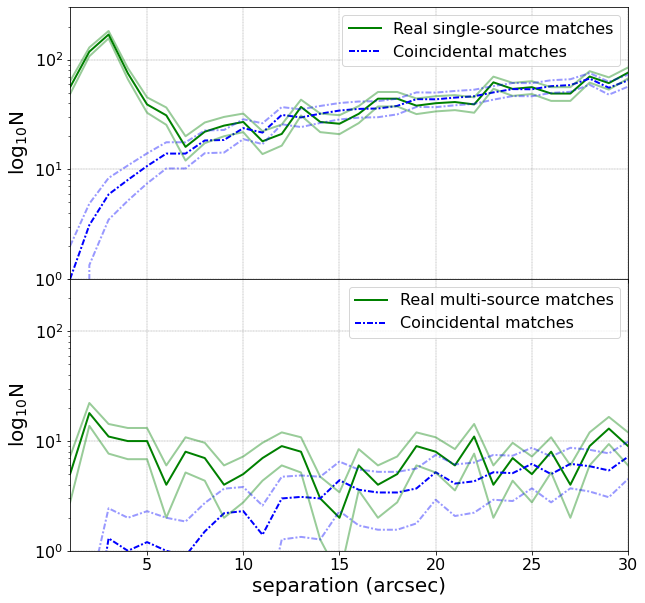}
    \caption{Statistical analysis of separation of single-source and multi-source FLASH sources versus WiggleZ galaxies in the W15 field, with separation in arcseconds along the x-axis, and the log of N - the number of matches - along the y-axis. Note that single-source and multi-source data are plotted separately, and that multi-source data contains one tenth of the amount of data. The real data consists of WIG\_SDSS and WIG\_SDSS\_EXT class data positionally cross-matched against FLASH objects in the W15 field, and is represented in both plots by the solid green line. The mean number of cross-matches of 10 datasets of positionally randomised data, shown with the broken blue line, is also plotted. Standard deviation for each set is included with lower opacity lines of respective style. We take 5\,arcsec to be the maximum limit of separation radius for genuine matches for both single and multi-source objects.}
    \label{fig:sources_sep}
\end{figure}

In order to probe the maximum radius for genuine associated matches between radio sources and host galaxies, a Monte Carlo test was done, and the result can be seen in Fig.\,\ref{fig:sources_sep}. This test was carried out on data in the W15 field, as it is the largest field containing the most uniform distribution of objects. It involved cross-matches between the 21461 WIG\_SDSS and WIG\_SDSS\_EXT class data in this field area with FLASH island objects, and comparing this against the averaged results of cross-matching 10 identically sized datasets of positionally randomised data. This was done separately both for FLASH islands consisting of a single component (simple sources), and for those determined to consist of more than one component (complex sources).

Out to $\sim$5\,arcseconds, the numbers of real matches for single source objects overwhelm any coincidental randomised data matches. This is taken to be the radius out to which cross-matches of radio sources and optical objects are genuine associated matches of optical galaxies and associated radio sources - for example, a galaxy with its synchrotron jet. Beyond 20\,arcsec, the data converge, and thus this is taken as an upper bound for object cross-matches that are `offset' - the radio source is not directly associated with the host galaxy, but is close enough to illuminate a \hi disk or CHM gas if present.

For multi-source objects on the other hand, the situation is more complex. Multi-component (complex) sources represent only 10\,percent of FLASH radio matches, so the data shows a larger degree of fluctuation due to the smaller number of matches. There is still enough difference between real data and randomised data to expect cross-matches out to 5\,arcsec for multi-source data to be genuine matches, however. We thus use the same associated and offset matching criteria for all radio source cross-matches, regardless of them being single-source or multi-source objects.

We assume that in the cases of WiggleZ-type optical objects which do not host a bright FLASH source, but are proximate to a FLASH source, that the FLASH source lies in the optical object's background, at a higher redshift than the optical object. This is based on the fact that bright FLASH sources (AGN) would usually lie at higher redshifts than galaxies not containing AGN - an example of this can be seen when comparing the average redshift of SF vs AGN host galaxies in Fig.\,\ref{fig:colour_magnitude_plots_5_20}. This is also supported by previous investigations of radio source redshifts \citep{Condon1998,deZotti2010}, as well as the results of other surveys of intervening absorbers \citep{Darling2011}. Indeed, while using ASKAP \citet{Sadler2020} successfully detected intervening absorption for 0.4\,<\,\textit{z}\,<\,1.0, and background sources were found to be either `in-band' (0.4\,<\,\textit{z}\,<\,1.0 but at a redshift above that of the optical galaxy) or ` background' (z\,>\,1.0).


\subsection{Optical-FLASH cross-matching}\label{flash_wigglez}

The FLASH continuum images have a typical rms noise of $\sim$0.1\,mJy (Yoon et al. \textit{in prep.}), but the catalogues produced by the automated pipeline contains some sources with flux densities as low as 0.2--0.3\,mJy\,beam$^{-1}$. 
Before cross-matching, a flux density cutoff of 0.5\,mJy\,beam$^{-1}$ was applied to ensure that only robustly-detected radio sources were matched. 

In total, the number of optical objects available for cross-matching against FLASH is 31663. These objects were cross-matched against the FLASH radio island catalogue to search for matches with an offset smaller than 20\,arcsec. 

The resulting matches were split into two classes based on the results of the Monte Carlo simulation presented in Fig.\,\ref{fig:sources_sep}. `Associated' radio sources were defined to be those with a radio-optical offset smaller than 5\,arcseconds, and `offset' sources those with a separation between 5 and 20\,arcseconds.

We can evaluate the completeness and reliability of our sample based off this figure. For single-source matches, out to 5\,arcseconds we see a clear separation between the population of real matches, and random matches. If we integrate beneath both of these curves, we get 401 matches for the real data, and 28.7 for the randomised data. Doing the same with the complex matches, we get 54 and 3.7 respectively. If we take the difference between genuine and random single source matches, and divide by the difference between total single and complex genuine and random source matches, we can calculate the completeness: (401\,-\,28.7)\,/\,(455\,-\,32.4) which results in 88\,percent. In terms of reliability, as we have found there to be 401 genuine matches and 28.7 randomised matches, we get a resulting 93\,percent.

The cross-matching process yielded 279 associated and 740 offset sources in total, corresponding to 3.2\,percent of the input optical sample. This is enough to perform analyses of the characteristics of both associated and offset cross-matches, involving comparison against the overall WiggleZ-like dataset comprising the W09, W15, and extended sample data.

For the production of the finalised catalogues, the cross-match process was repeated using the FLASH component catalogue, rather than the island catalogue for which this work used, in preparation for the next paper in this series.

\begin{figure}
    \centering
    \includegraphics[width=\linewidth]{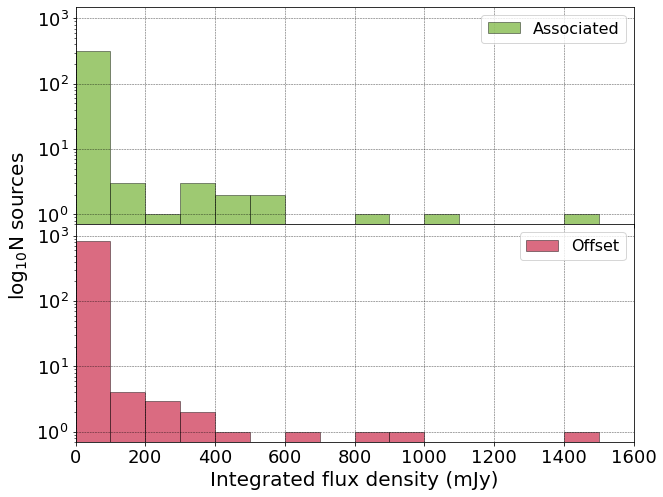}
    \caption{Histogram of integrated flux density (mJy) of radio source islands for associated (green) and offset (red) cross-matched FLASH-WiggleZ type objects. Note that the y-axis for each is logged. Few objects in both datasets possess very bright radio sources exceeding 100\,mJy.}
    \label{fig:flux_dens_hist}
\end{figure}

Fig.\,\ref{fig:flux_dens_hist} is a histogram of integrated flux density (mJy) for the associated and offset cross-matched objects. It is apparent for both populations that few radio sources that have been cross-matched exceed 100\,mJy in brightness. Indeed, 92\,percent of the associated objects and 95\,percent of the offset objects have integrated flux densities of 30\,mJy or less, which we treat as the threshold for a bright radio source.

Associated objects having a higher percentage of bright sources is unsurprising, since this kind of cross-match is representative of a galaxy containing a radio-loud AGN.

\section{AGN and Star-Forming Diagnostics}\label{bpt_whan}

\subsection{Diagnostic diagrams} \label{diagnostic_diagrams}

Comparison of degrees of ionisation of gases in galaxies allows for the classification of galaxy type. Various diagnostic plots have been developed, but one of the most widely used and recognised is the BPT \citep{Baldwin1981}. This uses line flux ratios of close pairs of emission lines in order to minimize the impact of reddening. For BPT diagrams, [OIII] ($\uplambda$4861) and H$\upbeta$ ($\uplambda$5007), and H$\upalpha$ ($\uplambda$6563) and [NII] ($\uplambda$6583) are used, plotted as log$_{10}$([OIII]/H$\upbeta$) against log$_{10}$([NII]/H$\upalpha$). A signal-to-noise ratio of at least 3 is required for lines to be included. Galaxies are then separated into three main classes: SF dominated, AGN dominated, and composite. This is only possible for objects with \textit{z}\,$\lesssim{0.48}$, as above this the [NII] and H$\upalpha$ line pair is redshifted into a spectral region with strong terrestrial airglow lines. 

WHAN diagrams are a second variety of diagnostic diagram, and were defined by \citet{Fernandes2010,Fernandes2011}. This plot was developed in order to reduce the number of spectral lines required in diagnostic diagrams, and uses only two [NII] and H$\upalpha$, in the form of the equivalent width of H$\upalpha$ against the log$_{10}$([NII]/H$\upalpha$) line ratio. These also tend to be two of the brightest spectral lines detectable in galactic spectra. This allows WHAN diagrams to potentially include a larger number of objects than BPT diagrams, while still allowing classification between galaxies containing star formation and AGN \citep{Salzer2005}. However, the [NII] and H$\upalpha$ lines are redshifted beyond detection for WiggleZ at \textit{z}\,$\gtrsim{0.48}$.

By comparing the BPT and WHAN diagrams for associated and offset object data, as well as taking an overall look at the behaviour demonstrated by all WiggleZ-type objects with \textit{z}\,$\lesssim{0.48}$, we can analyse the classifications of optical objects for each dataset, and whether this fits our expectations for the data.

\begin{figure}
    \centering
    \includegraphics[width=\linewidth]{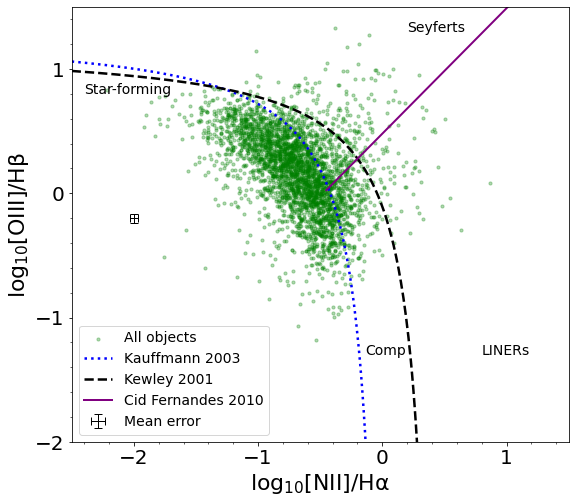}
    \caption{BPT diagnostic diagram of W09, W15, and extended sample field WiggleZ data pre radio source cross-matching along with extended sample data produced for this work, with redshift \textit{z}\,<\,0.48. 3391 systems are featured. Inclusion of \citet{Kewley2001} (black dashed line) and \citet{Kauffmann2003} (blue dotted line), means that classification of objects as AGN (above Kewley), star-forming (below Kauffmann), or composite (between) is possible. The diagonal solid purple line \citet{Fernandes2010} divides the AGN population between Seyfert and LINER, with the former above, and the latter below.}
    \label{fig:wigz_bpt}
\end{figure}

\begin{figure}
    \centering
    \includegraphics[width=\linewidth]{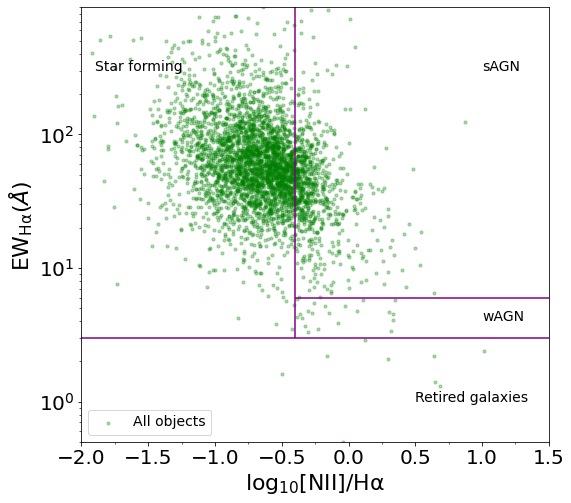}
    \caption{WHAN diagram of W09, W15, and extended sample field data with redshift \textit{z}\,<\,0.48. The 3766 systems of Fig.\,\ref{fig:wigz_bpt} have been featured. The equivalent width of the H$\upalpha$ line is plotted against the logged equivalent width ratio of NII and H$\upalpha$, as is standard for WHAN diagrams as laid out in \citep{Fernandes2010}. Horizontal and vertical diagnostic lines are from \citet{Fernandes2011}, separating star forming galaxies, strong AGN, weak AGN, and `passive galaxies' (none are included in this sample). Within the passive galaxies section, \citet{Fernandes2011} also includes `retired galaxies' or `fake AGN' (cyan), which have [NII] or H$\upalpha$\,<\,0.5\,Angstrom, on the right-hand side of the plot beneath the wAGN section.}
    \label{fig:wigz_whan}
\end{figure}

\subsection{Approach}\label{bpt_approach}

The method used in this work is to fit approximate single Gaussian curves using amplitude of peak and standard deviation, at the specific wavelengths of the spectral lines we are interested in. We do this to the optical spectra of all galaxies in our sample, which have been adjusted to their rest frame at redshift \textit{z}\,=\,0, using each object's observed spectroscopic redshift, and had its continuum either subtracted or normalised for line flux estimates or equivalent width estimates respectively. The line flux (LF) or equivalent width (EW) of the line is then derived from the fitted Gaussian.

Upon visual inspection of the Gaussians produced, an estimated <\,5\,percent are poor-quality fits, either being skewed from the spectral line position, or not well-fitted by a Gaussian. These introduce scatter into the diagnostic plots, as would be expected in an automated pipeline \citep{Drinkwater2018,Feuillet2023}. These would need to be excluded by hand, but even for rudimentary single Gaussian fitting, the vast majority were good matches. It is unlikely that their exclusion would have a great deal of impact on the appearance of resulting BPT and WHAN plots, other than reducing the scatter of the points plotted. However, the feasibility of checking multiple spectral lines for fit quality on thousands of points is not high. We note that after exclusion of detections below signal-to-noise ratio of 3, and along with the additional rejection criteria of the pipeline (see A2), the typical SNR of [OIII], H$\upbeta$, H$\upalpha$, and [NII] line fluxes for this sample are 15.4, 14.3, 33.9, and 18.5 respectively, and for the equivalent width of H$\upalpha$ was 18.6. These are still significantly higher than the threshold cutoff for detections. The pipeline identifies and measures emission lines that are not buried with noise, and is not likely to pick up spurious detections.

Synchrotron radiation can also contribute excess light in the continuum, which could be a concern in objects hosting a synchrotron source (e.g. AGN). For this work, it is unlikely that this effect will be important, as the diagnostic category will not impact on whether offset objects are tested for intervening 21\,cm absorption signatures.

When working with line fluxes, corrections for dust extinction and stellar absorption are sometimes put into place. While this allows for more accurate results, it is not strictly necessary to do so. We note for the reader that these corrections have not been made to the line flux data.

\subsection{Diagnostic diagrams for the WiggleZ and FLASH-matched objects }\label{flash_diagnostics}

\subsubsection{The WiggleZ optical sample}
Fig. \ref{fig:wigz_bpt} and \ref{fig:wigz_whan} show all the WiggleZ-type objects from W09, W15, and the extended field that pass the criteria outlined in sections \ref{diagnostic_diagrams} and \ref{bpt_approach}. From the BPT, it is apparent that WiggleZ-type objects as a whole exhibit predominantly star-forming behaviour, with the bulk lying beneath the star-formation line as developed by \citet{Kauffmann2003}. This is what we would expect, given that over 90\,percent of WiggleZ galaxies are star-forming \citep{Drinkwater2018}. 
A smaller number of objects lie between the Kauffmann line and the AGN line, described by \citet{Kewley2001}.

We also see a very similar result from the WHAN, where the bulk of these objects fall into the SF region, and far fewer are in either AGN category. 

How well do the BPT and WHAN results agree? We find that 92.8\,percent of WiggleZ-type objects classified as star-forming or AGN using the BPT are also found to be their respective classification by the WHAN. The remaining 7.2\,percent have a disagreement between categories. As we exclude composite objects when considering this, the objects where there is disagreement are likely to be objects on the BPT diagram which fall above the \citet{Kewley2001} line on the left-hand side of the plot, where they are classified as AGN here, and are likely then to fall into star-forming on the WHAN. As this number is reasonably low, and likely due to the difficulty in placing discrete boundaries on real-world phenomena, an amount of fuzziness in classification is not a serious concern. 

\subsubsection{The FLASH-optical matched sample}
\begin{figure}
    \centering
    \includegraphics[width=\linewidth]{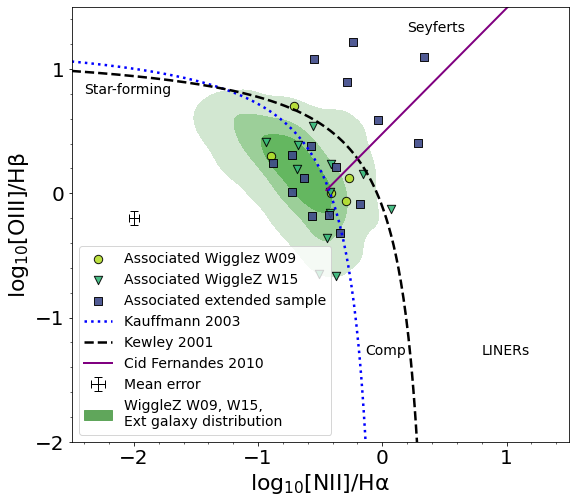}
    \caption{BPT diagnostic diagram of associated FLASH-WiggleZ/extended survey data with redshift \textit{z}\,<\,0.48. 33 systems are featured in total, including two SF galaxies which closely overlap. Diagnostic lines are as before in Fig.\,\ref{fig:wigz_bpt}, and the mean error of the objects is displayed. The contour is object density for all WiggleZ-type galaxies from Fig. 7, with a threshold excluding the 10 percent of objects least densely clustered in the population. Each contour contains a third of the WiggleZ-type galaxies above this threshold. The most common object classification is AGN, with 25.0\,percent of objects receiving this label, $\sim{}$5 times the rate of offset and unmatched WiggleZ-type objects.}
    \label{fig:flash_wigz_5_bpt}
\end{figure}

\begin{figure}
    \centering
    \includegraphics[width=\linewidth]{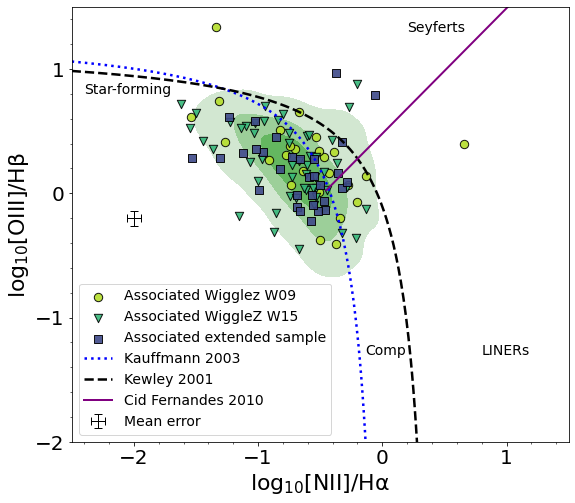}
    \caption{BPT diagnostic diagram of offset FLASH-WiggleZ/extended survey data with redshift \textit{z}\,<\,0.48. 107 systems are featured in total. Diagnostic lines, errors, and contours are as in Fig. \ref{fig:flash_wigz_5_bpt}. The most common object classification is star-forming, with 76\,percent of objects receiving this label.}
    \label{fig:flash_wigz_20_bpt}
\end{figure}

\begin{figure}
    \centering
    \includegraphics[width=\linewidth]{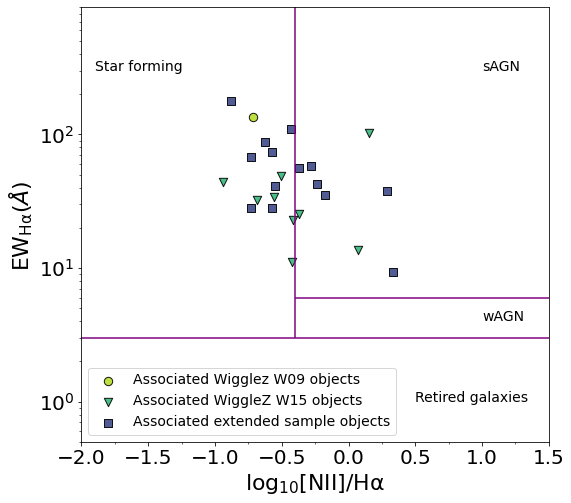}
    \caption{WHAN diagram of associated FLASH-WiggleZ/extended sample data with redshift \textit{z}\,<\,0.48. 24 systems are featured in total. The equivalent width of the H$\upalpha$ line is plotted against the logged EW ratio of NII and H$\upalpha$, as is standard for WHAN diagrams as laid out in \citep{Fernandes2010}. Diagnostic lines are as before in Fig.\,\ref{fig:wigz_whan}. It should again be noted that in this cross-matched sample, there are few wAGN, and no retired or passive galaxies. 37.5\,percent of objects receive the AGN designation, around twice the rate of the offset and unmatched WiggleZ-type objects.}
    \label{fig:flash_wigz_5_whan}
\end{figure}

\begin{figure}
    \centering
    \includegraphics[width=\linewidth]{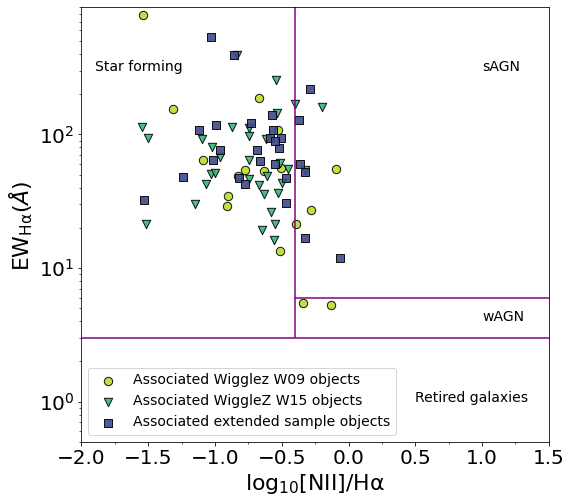}
    \caption{WHAN diagram of offset FLASH-WiggleZ/extended sample data with redshift \textit{z}\,<\,0.48. 107 systems are featured in total. Diagnostic lines are as before. It should be noted that in this cross-matched sample, there are no passive galaxies, and few wAGN or retired galaxies. 83.3\,percent of objects receive the star-forming designation.}
    \label{fig:flash_wigz_20_whan}
\end{figure}

We can now look at the BPT and WHAN diagrams of the small subset of WiggleZ-type objects that have been cross-matched with the FLASH catalogue, and the results for the much smaller sample of FLASH cross-matched objects are shown in 
Fig. \ref{fig:flash_wigz_5_bpt}, \ref{fig:flash_wigz_20_bpt}, \ref{fig:flash_wigz_5_whan}, and \ref{fig:flash_wigz_20_whan}. 

Unlike Fig.\,\ref{fig:wigz_bpt}, the BPT diagram for `associated' radio matches shown in Fig.\,\ref{fig:flash_wigz_5_bpt} has proportionally more objects in the AGN region of the plot. This is consistent with our expectation that many of the close radio matches are high-excitation radio AGN. There is a similar number of objects lying within the SF and composite sections, and these objects may correspond to galaxies in which a low-excitation AGN coexists with rapid ongoing star-formation. The WHAN diagram for associated data (Fig.\,\ref{fig:flash_wigz_5_whan}) shows a similar effect, with a greater number of AGN being detected for the associated objects compared to unmatched WiggleZ-type objects.

\begin{table}
\centering
\begin{tabular}{c|c|c|c|c|c|c}
    \hline \hline \multirow{2}{*}{$\mathrm{BPT}$}
     & \multicolumn{2}{c}{$\mathrm{SF}$} & \multicolumn{2}{c}{$\mathrm{AGN}$} & \multicolumn{2}{c}{$\mathrm{Comp}$}\\
     \cmidrule(lr){2-3} \cmidrule(lr){4-5} \cmidrule(lr){6-7}
     & N & \% & N & \% & N & \%\\
    \hline    
    Associated & 14 & 43.7\% & 8 & 25.0\% & 10 & 31.2\%\\
    Offset & 76 & 71.0\% & 5 & 4.7\% & 26 & 24.2\%\\
    WiggleZ-type & 2591 & 76.4\% & 181 & 5.3\% & 619 & 18.3\%\\
    \hline

    \hline \hline
    \multirow{2}{*}{$\mathrm{WHAN}$}
     & \multicolumn{2}{c}{$\mathrm{SF}$} & \multicolumn{2}{c}{$\mathrm{AGN}$} & \multicolumn{2}{c}{$\mathrm{Retired}$}\\
     \cmidrule(lr){2-3} \cmidrule(lr){4-5} \cmidrule(lr){6-7}
     & N & \% & N & \% & N & \%\\
    \hline
    Associated & 15 & 62.5\% & 9 & 37.5\% & 0 & 0.0\%\\
    Offset & 65 & 83.3\% & 13 & 16.7\% & 0 & 0.0\%\\
    WiggleZ-type & 3028 & 80.4\% & 729 & 19.6\% & 9 & 0.2\%\\
    \hline
\end{tabular}
\caption{Breakdown of object classifications for associated and offset objects, as well as all WiggleZ-type objects (pre cross-matching), as determined by BPT and WHAN diagnostics. The number of objects assigned a classification, as well as the percentage this represents, is provided. There are 32, 107, and 3391 total associated, offset, and unmatched WiggleZ-type objects featured in the BPT, and 33, 78, and 3766 total associated, offset, and unmatched WiggleZ-type objects featured in the WHAN respectively. It can be seen that the offset data classification more similarly matches that of all WiggleZ type objects compared to the associated.}
\label{tab:bpt_whan_classifications}
\end{table}

On the other hand, the offset cross-matched data (Fig. \ref{fig:flash_wigz_20_bpt} and \ref{fig:flash_wigz_20_whan}) show quite different distributions from that of the associated data. In the BPT, most objects fall below the Kauffmann curve, being classified as star-forming galaxies. Again, this behaviour is echoed in the WHAN diagram, and objects in the sAGN section lie on the edge of the section closest to SF, rather than being scattered throughout that section.

In all three WHAN diagrams, very few objects fall into the wAGN section, which \citet{Fernandes2011} describes as `weak AGN', such as LINERs, which generally have low levels of UV activity \citep{Tadhunter2002} and so would be excluded from the WiggleZ optical sample. 

The results from the BPT and WHAN diagrams in this section are summarized numerically in Table \ref{tab:bpt_whan_classifications}. Here, we can directly compare the breakdown of classifications for each of the associated, offset, and WiggleZ-type object datasets, beyond visual inspection of the plots. The similarity in classification rate of the offset and overall dataset of WiggleZ objects are consistent between the BPT and WHAN, showing that most of the offset objects do not contain AGN, and are likely to be UV-bright star-forming galaxies that happen to be close on the sky to a bright radio source.

On the other hand, the associated objects clearly exhibit a different kind of behaviour and are two to five times as likely as the offset or WiggleZ-type objects to contain an AGN.


\section{Discussion}\label{discussion}

\subsection{The radio detection rate for WiggleZ galaxies}

The WiggleZ galaxies are selected for UV-brightness, but there is no direct bias for or against radio loudness with these objects. From a cross-match of the WiggleZ/extended sample with the FLASH radio catalogue, we found that fewer than 1\,percent of the WiggleZ galaxies (279/31663) are `radio-loud' (i.e. associated with a radio source brighter than 0.5\,mJy\,beam$^{-1}$). 
 
A further 2.3\,percent of WiggleZ galaxies lie between 5 and 20\,arcsec on the sky from a FLASH radio source. These galaxies are unlikely to be physically associated with the radio source, but if the radio source lies at higher redshift then it can be used as a probe to search for \hi absorption at the redshift of the WiggleZ galaxy. 
 
The two catalogues presented in the Appendix provide the tools needed for the \hi absorption that will be the focus of the second paper in this series. At the median WiggleZ redshift of z\,=\,0.6, the angular scale on the sky is 6.76\,kpc\,arcsec$^{-1}$. A 5\,--\,20\,arcsec offset therefore corresponds to an impact parameter of around 30\,-\,120\,kpc and typically probes the CGM, while an offset smaller than 5\,arcsec probes gas within the galaxies themselves.

\begin{table}
\centering
\begin{tabular}{l|r|r|r}
    \hline \hline
    Field & N\textsubscript{associated} & N\textsubscript{offset} & N\textsubscript{$\upSigma$(field)}\\
    \hline    
    W09 & 29 & 158 & 5813\\
    W15 & 133 & 379 & 18388\\
    Extended Sample & 117 & 203 & 7462\\
    \hline
    Total & 279 & 740 & 31663\\
    \hline
\end{tabular}
\caption{Table of field, cross-match type, and number of objects cross-matched against FLASH, as well as total number of optical objects in field pre positional cross-matching. The small size of the fraction of objects in each field that can be positionally cross-matched can be seen.}
\label{tab:cross_match_numbers}
\end{table}

\subsection{Interpretation of the diagnostic diagrams} 

The BPT and WHAN diagnostic plots allow us to analyse the behaviour of WiggleZ-type galaxies, with the BPTs showing them to fall into star-forming, AGN, or composite categories, and the WHAN diagrams showing that galaxies possessing AGN overwhelmingly lie in the `strong AGN' category (an analogue for Seyferts proposed in \citet{Fernandes2011}, where WHAN diagrams were first being developed). This is not surprising, considering the well observed link between the UV brightness of objects, and their star-forming and/or AGN behaviour \citep{Tadhunter2002}.

Using the extreme star formation diagnostic developed in \citet{Kewley2001}, \citet{Drinkwater2010}, which combines both SF and composite categories, finds that 90\,percent of WiggleZ galaxies plotted on a BPT diagram are star-forming. This shows that their selection criteria have succeeded in selecting for UV-bright star-forming galaxies while managing to largely exclude AGN. We similarly have found that 94.7\,percent of our extended WiggleZ-type catalogue are classed as star-forming when using the same extreme star formation envelope. The WHAN diagram also demonstrates that the large majority of objects in the extended WiggleZ-type catalogue are star-forming, however at a lower rate (80.4\,percent). This is likely due to the difference in the two diagnostic plots, where WHAN takes into account less lines, and the positioning of the different diagnostic regions. 

 Considering the link between AGN, star-formation, and the abundance of \hi found in galaxies displaying these types of behaviour \citep{Gereb2015,Maccagni2014,Rhee2023} implies that UV brightness and \hi content should be connected. Pairing the FLASH survey with WiggleZ-type galaxies therefore provides an excellent opportunity to search for 21\,cm absorption in a poorly-studied redshift range \citep{Dutta2022}, without relying on radio selection bias, which is a common hurdle in 21\,cm studies, where 21\,cm signals tend to be found more commonly at lower redshifts (\textit{z}\,<\,0.4), which is thought to be caused by more extreme UV-luminosities of AGN, or the powerful radio emissions of bright radio sources associated with AGN, which tend to be more common at higher redshifts \citep{Minchin2003,Aditya2024}. 
 Our selection of UV-bright star-forming objects also allows us to test how the UV luminosity of non-AGN radio-source host galaxies affects the detection rate in associated 21\,cm absorption searches \citep{Curran2012}.


\subsection{Differences in BPT distribution between associated and offset cross-matched galaxies}\label{bpt_pattern}
 
In Section \ref{flash_diagnostics}, BPT diagrams for the associated and offset cross-matches with FLASH were produced. With regard to the BPT diagrams, it is interesting to compare the difference in apparent behaviour obtained for associated data (Fig.\,\ref{fig:flash_wigz_5_bpt}) versus those obtained for offset data (Fig.\,\ref{fig:flash_wigz_20_bpt}), as well as the numerical values as presented in Table \ref{tab:bpt_whan_classifications}. In the case of associated data - though the sample size is limited, it appears an associated object is 5 times more likely to be classified as an AGN host.

However, in the case of offset galaxies, the distribution is quite different. In this case, there are many more galaxies classed as star-forming, far exceeding the combined AGN and composite classifications. It seems to suggest that when searching around a radio source at an extended distance for non-associated UV-bright objects, you are more likely to detect purely star forming objects than any other category, and this result is consistent regardless of looking specifically at W09, W15, or the extended sample data.

Considering the observed connection between \hi reservoirs and star-formation, as well as the offset category of FLASH-WiggleZ cross-matched objects being the ideal targets for intervening 21\,cm absorption probing, this is a positive result for our work.


\subsection{Optical colours and redshift}\label{colour_z}

 \begin{figure}
    \centering
    \includegraphics[width=\linewidth]{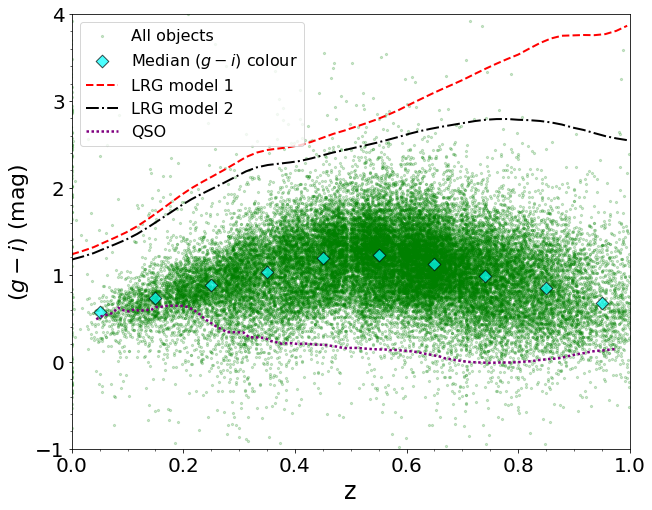}
    \caption{The (\textit{g\,-\,i}) colour versus redshift for all objects meeting the required selection criteria in the W09, W15, and extended sample fields. The median (\textit{g\,-\,i}) magnitudes for each redshift bin (cyan diamonds) are overlaid over the scatter, positioned at the centre of each bin (i.e. at \textit{z}\,=\,0.05 for 0\,<\,\textit{z}\,<\,0.1). The greater the (\textit{g\,-\,i}) value, the redder the object. The distribution is fairly symmetric, with its peak in reddest population around \textit{z}\,$\sim{0.55}$. LRG passive evolution tracks from \citet{Wake2006,Ching2017} (red dashed and black dot-dashed lines) and a QSO track \citet{Schneider2005,Croom2009} (purple dotted line) are included. 90\,percent of WiggleZ-type objects are star-forming galaxies \citep{Drinkwater2010} and, as we expect, this population does not follow the evolutionary track of either LRGs or QSOs.}
    \label{fig:wigz_colour_mag}
\end{figure}

\begin{figure*}
\begin{multicols}{2}
    \includegraphics[width=\linewidth]{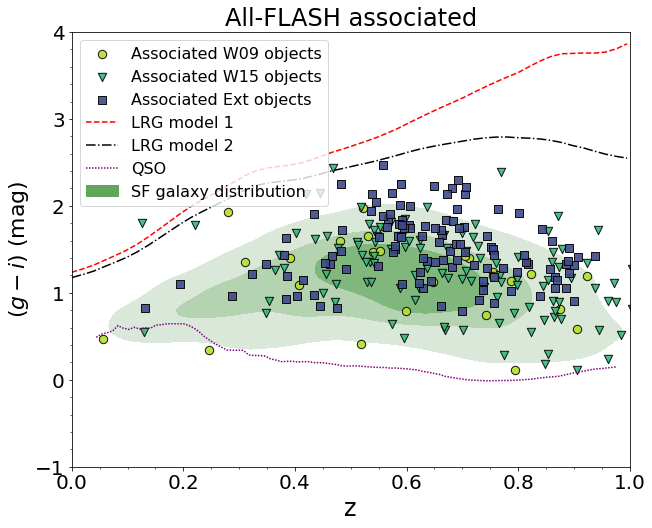}
    \par
    \includegraphics[width=\linewidth]{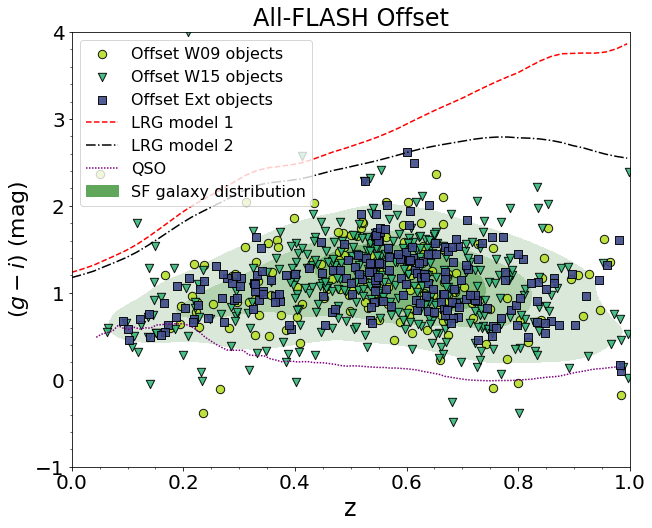}
\end{multicols}

\caption{(\textit{g\,-\,i}) colour versus redshift diagrams for associated and offset data, including evolutionary tracks from Fig.\,\ref{fig:wigz_colour_mag}. \textit{Left} contains all associated FLASH-WiggleZ type object cross-matches with the different fields indicated by different object markers (W09 - green circles, W15 - teal triangles, and extended sample - purple squares), while \textit{right} contains all offset cross-matches with identical convention for marker types. The included contour is SF object density from Fig.\,\ref{fig:wigz_colour_mag} with a threshold excluding the 15\,percent of objects least densely clustered in the population, and each contour containing a third of the SF population above this threshold. The increased redness and redshift of associated objects compared to star-forming WiggleZ-type objects, as well as offset objects, can be seen.}
\label{fig:colour_magnitude_plots_5_20}
\end{figure*}

In galaxy colour-magnitude diagrams, a bluer colour corresponds to actively star-forming galaxies, red to `dead', inactive galaxies, and there is a third population of transitional `green valley' galaxies between them, where galaxies that have been stripped of gas through internal processes, or mergers lie \citep{Fabian1999,Martin2007,Wyder2007}. Fig. \ref{fig:wigz_colour_mag}, and \ref{fig:colour_magnitude_plots_5_20} show the relationship between (\textit{g\,-\,i}) colour magnitude, and redshift for the whole WiggleZ-type population, and for our two FLASH cross-matched catalogues. The larger the value of (\textit{g\,-\,i}), the redder the object. We can then study how this changes with redshift, for our different populations of objects.

Fig.\,\ref{fig:wigz_colour_mag} shows the entire WiggleZ-type data sample in the same format as fig.\,14 of \citet{Ching2017}. The evolutionary tracks included in Fig.\,\ref{fig:wigz_colour_mag} are also the same, including two `luminous red galaxy' (LRG) models using data from \citet{Wake2006} (dashed red line and dash-dotted black line) to trace out `low excitation radio galaxy' (LERG) evolution, and a single `quasi-stellar object' (QSO) track (dotted purple) \citep{Schneider2005,Croom2009}.

The WiggleZ-type galaxies lie between the regions of LRGs that exhibit passive evolution, and QSOs which maintain a consistent bluer colour. As the WiggleZ survey contains 90\,percent star-forming galaxies \citep{Drinkwater2010}, this region on the (\textit{g\,-\,i}) vs \textit{z} plot contains star-forming galaxies. The median (\textit{g\,-\,i}) colour changes with redshift, reaching a maximum at \textit{z}\,$\approx{0.55}$, before turning over and returning to a similar value at \textit{z}\,=\,1.0 as at \textit{z}\,=\,0.

In Fig.\,\ref{fig:colour_magnitude_plots_5_20}, the median redshift for associated objects is \textit{z}\,=\,0.65, whereas for offset objects it is \textit{z}\,=\,0.57. This difference in median redshift for the two radio cross-match samples may be due to the higher AGN fraction in the associated sample, since radio-loud AGN are more common at higher-redshift \citep[e.g.][]{pracy16}. On the other hand, we have shown that the offset objects are not generally hosts of AGN, and they have a much more even redshift spread than the associated objects.

The contours in Fig.\,\ref{fig:colour_magnitude_plots_5_20} show the star-forming population of WiggleZ-type galaxies from Fig.\,\ref{fig:wigz_colour_mag}, and are useful for comparing the associated and offset objects. The SF contours are based on density, and have a threshold cutoff applied, so the least clustered 15\,percent of SF objects from Fig.\,\ref{fig:wigz_colour_mag} are excluded. This reduces the extent of the contour plot and excludes any objects that aren't SF galaxies. The contours are also split into three levels based on object density, each containing a third of the population.

The associated objects do not fit the SF distribution well. They tend to have redder colours and higher redshift, which is expected, as BPT and WHAN both classified the associated objects as very likely to contain an AGN. Contrastingly, the offset population fits this distribution well, not only in terms of the shape of its distribution, but also its position and area of highest density. This is again confirmation that the offset objects are star-forming, like the overall population of WiggleZ-type objects.

We also conclude from this that the offset objects are otherwise typical SF WiggleZ-type objects that happen to be close to bright FLASH sources, and that it is probable that any \hi characteristics that these offset objects have would be similar to those of the overall WiggleZ-type dataset.


\subsection{Single and multiple-component radio sources}\label{source_component_type}

It is important to be aware that radio sources can possess different structures. Although the majority of radio sources are simple compact sources, it is still possible that larger, extended, and/or multi-component radio sources can be either associated with galaxies, or occur in their background. If this is the case for a radio source, ASKAP surveys flag this in data releases. Therefore, the data used in this work also has a flag indicating the number of components that the radio source is comprised of. The \verb/Selavy/ algorithm \citep{Whiting2012b} is responsible for ASKAP's ability to compare sources in 3D space, and evaluates the likelihood of their apparent 2D relationships being true in 3D space.

From the two FLASH cross-matched populations (associated and offset), 
it appears that neither type of cross-match is significantly more likely to occur with a multi-component source, occurring at a rate of $\sim$10\,percent (see Table \ref{multi_source_matches}). This also matches with Fig.\,\ref{fig:sources_sep}, where complex sources account for roughly 10 percent of total radio source matches.

Only offset matches with the W15 field exceed this $\sim$10\,percent rate, with 65 out of 379 objects being complex source matches. Given the number of objects in W15 being the greatest for all fields, this is quite interesting. Considering that the other two offset cross-matches do not exhibit this increased rate, it would be interesting to understand whether this is purely a random spike in complex radio sources in this field for offset objects, or if there is an underlying process. When checked against older observations of FLASH fields in this region, the elevated rate of detections remains.

It is shown in fig.\,18 of \citet{Drinkwater2010} that a high proportion of WiggleZ galaxies are in some way interacting with others, and they also discuss that there is a degree of clustering between the UV-bright WiggleZ galaxies, though much lower than what would be observed in LRGs. This would be expected to be seen throughout the WiggleZ selection, and is likely not the cause of the apparent increase in complex radio sources in the W15 field.

As shown in Fig.\,\ref{fig:redshift_hist}, all three fields involved in this work (W09, W15, Ext) probe to similar redshifts, and have similar distributions of galaxies across redshift - particularly the two WiggleZ fields, W09, and W15 - so it would be expected that a fluctuation to this degree of offset cross-matches would be seen in W09 as well as W15, especially considering the lower number of galaxies contained within the W09 field.

As part of the WiggleZ survey design is to probe over-densities in space caused by BAO, there is a possibility it could be a real physical effect due to more over-dense regions of cosmic structure coinciding with the W15 field than with other fields. This is difficult to tell from comparing to fig. 14 and 16 of the GAMA DR1 paper \citep{Driver2011} which carried out spectroscopic redshift experiments and charted cosmic structure in many of the fields covered by the full WiggleZ survey, in particular the 9h and 15h regions which our W09 and W15 fields cover parts of. A complete investigation of potential causes is outside the scope of this work, but would be of interest in our second paper in this series.

\begin{table}
\centering
\begin{tabular}{l|l|r|r|r}
    \hline \hline
     & Field & N\textsubscript{complex} & N\textsubscript{total} & \% multi-source \\
     & & & & cross-matches\\
    \hline    
    Associated & W09 & 3 & 29 & 10.3 \\
     & W15 & 14 & 133 & 10.5 \\
     & Ext. sample & 14 & 117 & 12.0\\
     \hline
    Offset & W09 & 18 & 158 & 11.4 \\
     & W15 & 65 & 379 & 17.2 \\
     & Ext. sample & 19 & 203 & 9.3\\
    \hline
\end{tabular}
\caption{Number and percentage of multi-component FLASH radio objects that each field contains, as well as total radio cross-matches for each field, with associated and offset cross-matches displayed separately.}
\label{multi_source_matches}
\end{table}


\subsection{Distribution of galaxies with redshift}

\begin{figure}
    \centering
    \includegraphics[width=\linewidth]{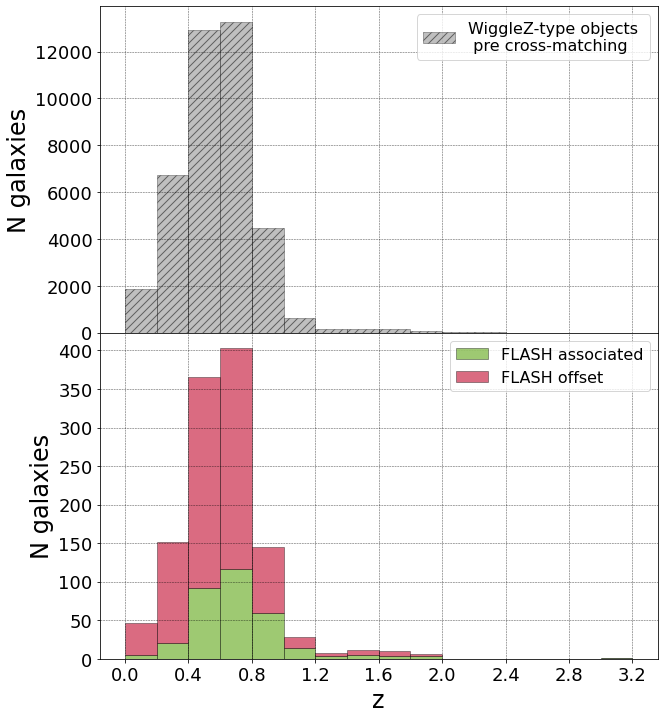}
    \caption{Distribution of redshifts for objects in total WiggleZ survey with WIG\_SDSS class (top) and the two datasets of cross-matched objects stacked on top of one another (bottom). Associated data is green, and offset data is red. Note the shared x-axis and different scales on the y-axes, as the number of cross-matched objects cumulatively is far smaller than the entire WiggleZ-type dataset. Due to small number statistics, any conclusions from \textit{z}\,>\,1.2 should be discounted.}
    \label{fig:z_hist}
\end{figure}

As can be seen from Fig.\,\ref{fig:z_hist}, the complete set of WiggleZ-type galaxies (grey hatched bins) possess a distribution where the two most populated bins occur at 0.4\,<\,\textit{z}\,<\,0.6, and 0.6\,<\,\textit{z}\,<\,0.8, and reduce steeply to either side. This is not unexpected, as the median WiggleZ redshift is \textit{z}\,=\,0.6 \citep{Drinkwater2018}. For the specific regions used in this paper, median redshifts are \textit{z}\,=\,0.57, \textit{z}\,=\,0.58, and \textit{z}\,=\,0.59 for W09, W15, and the extended sample respectively. The overall median for all objects considered in this work is \textit{z}\,=\,0.58.

The second histogram of Figure \ref{fig:z_hist} contains stacked bins for WiggleZ-type objects that are associated with a FLASH source, and for objects that are offset from a FLASH source. As these are separate populations, there is no duplication of a WiggleZ-type object in this process. The resulting stacked histogram allows us to fully understand the redshift distribution of cross-matched objects, and understand the contribution from the different cross-matches visually.

From the similarity in distribution in terms of the median peak, and the sharp decline to either side of this peak, it is reasonable to conclude that the WiggleZ-type objects we have studied in W09, W15, and the extended sample are representative of the whole WiggleZ population, possessing similar redshift distributions, median redshifts, and colour magnitude against redshift distributions.


\subsection{Mass-excitation diagnostic}\label{mex}

The WiggleZ survey additionally contains stellar masses for galaxies with \textit{z}\,>\,0.3; see \citet{Drinkwater2018} for detailed method. Galaxies in all three of our fields may contain stellar mass information, as all three fields do contain WiggleZ galaxies. Medians for all galaxies containing stellar mass estimates in the extended sample, for associated galaxies containing stellar mass, and for offset galaxies containing stellar mass, are 9.94, 10.805, and 10.34 log\,(M$_{*}$/M$_{\odot}$) respectively. Presence of a radio-loud AGN is known to correlate with a higher stellar mass, so it is unsurprising that our associated sample would have the highest median mass compared to offset galaxies or our overall extended catalogue \citep{Best2005a,Wang2008}.

Stellar mass can be used against the [OIII]/H$\upbeta$ line ratio to produce an MEx plot \citep{Juneau2014}. This plot allows for the separation of SF galaxies from those containing AGN, which is useful for the WiggleZ sample, where the [NII] and H$\upalpha$ lines are no longer detectable for \textit{z}\,>\,0.48.

\begin{figure}
    \centering
    \includegraphics[width=\linewidth]{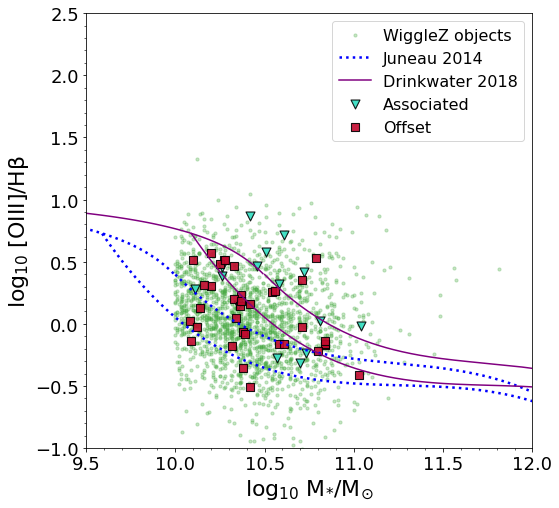}
    \caption{Plot of [OIII]/H$\upbeta$ line ratio against stellar mass for WiggleZ objects with \textit{z} $\geq$ 0.3 (green dots). Stellar mass cutoff of (log\,M$_{*}$/M$_{\odot}$)\,<\,10 included as in fig.\,5 of \citet{Drinkwater2018}. Demarcation lines from \citet{Juneau2014} separating likelihoods for AGN are included for low redshift galaxies (dotted blue lines), as well as the shifted demarcation line produced specifically for the line luminosity detection limit of WiggleZ galaxies from \citet{Drinkwater2018} (solid purple lines). Associated and offset galaxies that contain WiggleZ stellar mass information as well as [OIII]/H$\upbeta$ information are plotted as teal triangles and red squares respectively.}
    \label{fig:mex}
\end{figure}

Fig.\,\ref{fig:mex} is an MEx plot for all objects in this work that contain spectroscopic redshift data from the WiggleZ survey (W09, W15, and some Ext objects with WiggleZ spectra), which also are classified as `high stellar mass' galaxies (10\,<\,log\,(M$_{*}$/M$_{\odot}$)\,<\,12), before any cross-matching has been undergone, and had \textit{z}\,$\geq$\,0.3, allowing them to be plotted in the same manner as in fig.\,5 of \citet{Drinkwater2018}. Associated and offset galaxies that have WiggleZ stellar mass and [OIII]/H$\upbeta$ information are plotted on top of this as teal triangles and red squares respectively. The demarcation lines from \citet{Juneau2014} (dotted blue lines) allow the sorting of low redshift objects by probability of containing AGN, with objects between the two lines having P$_{\mathrm{AGN}}$\,$\sim{0.4}$, and above the upper line having P$_{\mathrm{AGN}}$\,$\sim{0.7}$. However, for WiggleZ galaxies, this criteria does not as effectively sort galaxies into these classes.

As in \citet{Drinkwater2018}, we have included an adjusted version of the demarcation lines from \citet{Juneau2014}, which were shifted to take into account the luminosity of the WiggleZ objects. These are indicated with the solid purple lines. From this, we see again that the majority of WiggleZ objects are very likely to exhibit SF behaviour, rather than AGN. We also see that while offset objects are most likely to be classed as star-forming, and least as AGN, 7 out of 13 associated objects are likely to be AGN, just over 50\,percent.

We have applied a stellar mass cut where objects with log\,(M$_{*}$/M$_{\odot}$)\,<\,10 are excluded from the plot. This is explained in \citet{Drinkwater2018}, as higher mass WiggleZ galaxies were found to have significantly lower metallicities than regular emission line galaxies of similar mass. This does not hold for lower mass WiggleZ galaxies, where their metallicity is on par with regular emission line galaxies of similar mass, so these objects are excluded from the plot.

As can be seen in Fig.\,\ref{fig:mex}, when comparing the WiggleZ objects against the shifted demarcation line from \citet{Drinkwater2018}, the vast majority of both WiggleZ-type and offset objects are predicted to be star-forming, as most lie beneath the lowermost solid purple line. This agrees strongly with the conclusions we have drawn from BPT, WHAN and (\textit{g\,-\,i}) colour evolution plots for the overall sample of WiggleZ-type galaxies - the UV-brightness of most of these galaxies is driven by extremely high rates of star-formation.

\section{Conclusion}\label{conclusion}

We have produced two catalogues of optically unbiased, UV-bright galaxies designed to probe associated and intervening \hi absorption at intermediate redshifts (0.4\,<\,\textit{z}\,<\,1.0) have been produced. Our `associated' catalogue contains 279 objects and is designed for a search for associated \hi absorption. Our `offset' catalogue contains 740 objects, and has been designed for intervening \hi absorption studies involving spectral stacking. We have investigated the properties of these catalogues in various ways, and present a summary of our findings below:

\begin{enumerate}[i.]
\item
Two subsamples of radio cross-matched UV-bright star-forming galaxies have been produced - the `associated' catalogue for the investigation of associated \hi absorption and the `offset' catalogue for \hi absorption stacking for intermediate redshifts. These galaxies are also poised for investigation of the effect of high rates of star formation on neutral gas.

\item
BPT, diagnostic diagrams classify WiggleZ-type objects pre cross-matching as being majority (76.4\,percent) star forming. 
For the radio-matched sub-samples, only 43.7\,percent of the associated objects are classified as SF, while 71.0\,percent of offset objects are classified as SF galaxies. In comparison, WiggleZ-type objects pre cross-matching and offset objects are found to be 5.3 and 4.7\,percent AGN, while associated objects are 25.0\%, 5 times higher. This is as we expect - the cross-matching criteria of the associated objects searches for galaxies hosting bright radio sources, such as AGN. On the other hand, the offset objects are WiggleZ-type objects selected to be in the vicinity of bright radio objects, but not hosting them. While there is some fluctuation between rates of galaxies classified as SF/AGN between BPT, WHAN and MEx diagrams, there is a good degree of agreement for classification trends for associated, offset, and unmatched WiggleZ-type objects.

\item
When we look at the evolution of (\textit{g\,-\,i}) colour over redshift for all WiggleZ-type objects, we see that they lie between the passive evolutionary tracks of LRGs, and the flat QSO track. We identify this region as being the typical space occupied by SF galaxies. When we then compare the locations occupied by associated and offset objects, we see that while the offset objects lie mainly in the SF region, the associated objects are both redder, and lie at higher redshifts. We expect these findings for both populations - associated objects host AGN, while offset objects are SF galaxies.

\item
The combined redshift distribution of associated and offset objects when compared to one another is very similar, and we assume from this, along with our findings from the BPT and WHAN diagnostic diagrams, and (\textit{g\,-\,i}) colour vs \textit{z} distribution, that when considered together, the population of associated and offset objects are representative of the WiggleZ-type object catalogue in general. It may be possible to draw conclusions on the \hi content of unmatched WiggleZ-type galaxies overall.

\item
The fraction of multi-component FLASH cross-matches is similar for associated and offset objects in the three different fields is similar, at $\sim$10\,percent, which also matches the rate of complex radio sources in the FLASH catalogue overall. Only the offset W15 field has a significantly higher fraction of multi-component sources, at $\sim$17\,percent. As the W15 field is the field containing the most objects, this is particularly interesting. It is not known whether this is a random fluctuation in the matches, or if there is an underlying physical cause. From initial checks against previous FLASH observations of fields in this region, the effect seems real, and will be investigated in the second paper in this series.

\end{enumerate}

\noindent
Detections of the presence of \hi between 0.4\,<\,\textit{z}\,<\,1.0 and beyond have until now required emission-line spectral stacking, averaging over a number of rest-frame spectra to produce a median spectrum, in order to reveal a signal that may exist otherwise buried in noise \citep{Chowdhury2021,Khandai2011}, or the production of intensity maps \citep{Masui2013} indicating the difficulty of detection of this signal beyond the local Universe. This has left attempts to accurately understand the interplay between AGN feedback, star formation, \hix, and galactic evolution incomplete \citep{Lewis2003}. Recently, FLASH has demonstrated its capabilities with the detection of an extremely strong associated absorber towards a bright compact radio galaxy lying at \textit{z}\,=\,0.851 \citep{Aditya2024}, as well as the discovery of two associated absorbers at lower redshifts (\textit{z}\,=\,0.522 and \textit{z}\,=\,0.563) in massive (M\,>\,$10^{11}$\,$\mathrm{M_{\odot}}$) GAMA galaxies, both of which also have very strong \hi absorption signals \citep{Su2022}, so it may be increasingly possible to make associated \hi absorption observations at intermediate redshifts in the Square Kilometre Array era.

In the second paper in this series, we will be looking in depth at the FLASH radio spectra of the associated and offset objects found in this work, including seeking out evidence of associated 21\,cm absorption around the objects with the brightest radio continua. It should still be noted that due to the intrinsic faintness of 21\,cm absorption, stacking will likely be required to confirm the presence of \hi absorption signals from spectra of objects found in this work, especially for objects not considered `bright' in this context (<\,30\,mJy). However, the study of 21\,cm absorption in this redshift range, and potential discovery of other bright absorbers, will contribute to an improved picture of the distribution of \hi throughout the Universe at this fascinating period in its history.


\section*{Acknowledgements}
SLE thanks STFC for support through a STFC doctoral scholarship. 

This scientific work uses data obtained from Inyarrimanha Ilgari Bundara / the Murchison Radio-astronomy Observatory. We acknowledge the Wajarri Yamaji People as the Traditional Owners and native title holders of the Observatory site. CSIRO’s ASKAP radio telescope is part of the Australia Telescope National Facility (https://ror.org/05qajvd42). Operation of ASKAP is funded by the Australian Government with support from the National Collaborative Research Infrastructure Strategy. ASKAP uses the resources of the Pawsey Supercomputing Research Centre. Establishment of ASKAP, Inyarrimanha Ilgari Bundara, the CSIRO Murchison Radio-astronomy Observatory and the Pawsey Supercomputing Research Centre are initiatives of the Australian Government, with support from the Government of Western Australia and the Science and Industry Endowment Fund.

Parts of this research were supported by the Australian Research Council Centre of Excellence for All Sky Astrophysics in 3 Dimensions (ASTRO 3D), through project number CE170100013.


This work was supported by the Global-LAMP Program of the National Research Foundation of Korea (NRF) grant funded by the Ministry of Education (No. RS-2023-00301976).


This research has made use of "Aladin Sky Atlas" developed at CDS, Strasbourg Observatory, France \citep{Bonnarel2000}, and the "Specutils" Python package, Nicholas Earl (2024) “astropy/specutils: v1.13.0”. Zenodo. doi: 10.5281/zenodo.10681408.


This research is based on observations made with the Galaxy Evolution Explorer, obtained from the MAST data archive at the Space Telescope Science Institute, which is operated by the Association of Universities for Research in Astronomy, Inc., under NASA contract NAS 5–26555.


Funding for SDSS-III has been provided by the Alfred P. Sloan Foundation, the Participating Institutions, the National Science Foundation, and the U.S. Department of Energy Office of Science. The SDSS-III website is http://www.sdss3.org/.

SDSS-III is managed by the Astrophysical Research Consortium for the Participating Institutions of the SDSS-III Collaboration including the University of Arizona, the Brazilian Participation Group, Brookhaven National Laboratory, Carnegie Mellon University, University of Florida, the French Participation Group, the German Participation Group, Harvard University, the Instituto de Astrofisica de Canarias, the Michigan State/Notre Dame/JINA Participation Group, Johns Hopkins University, Lawrence Berkeley National Laboratory, Max Planck Institute for Astrophysics, Max Planck Institute for Extraterrestrial Physics, New Mexico State University, New York University, Ohio State University, Pennsylvania State University, University of Portsmouth, Princeton University, the Spanish Participation Group, University of Tokyo, University of Utah, Vanderbilt University, University of Virginia, University of Washington, and Yale University.


Funding for the Sloan Digital Sky Survey IV has been provided by the Alfred P. Sloan Foundation, the U.S. Department of Energy Office of Science, and the Participating Institutions. SDSS acknowledges support and resources from the Center for High-Performance Computing at the University of Utah. The SDSS website is www.sdss4.org.

SDSS is managed by the Astrophysical Research Consortium for the Participating Institutions of the SDSS Collaboration including the Brazilian Participation Group, the Carnegie Institution for Science, Carnegie Mellon University, Center for Astrophysics | Harvard \& Smithsonian (CfA), the Chilean Participation Group, the French Participation Group, Instituto de Astrofísica de Canarias, The Johns Hopkins University, Kavli Institute for the Physics and Mathematics of the Universe (IPMU) / University of Tokyo, the Korean Participation Group, Lawrence Berkeley National Laboratory, Leibniz Institut für Astrophysik Potsdam (AIP), Max-Planck-Institut für Astronomie (MPIA Heidelberg), Max-Planck-Institut für Astrophysik (MPA Garching), Max-Planck-Institut für Extraterrestrische Physik (MPE), National Astronomical Observatories of China, New Mexico State University, New York University, University of Notre Dame, Observatório Nacional / MCTI, The Ohio State University, Pennsylvania State University, Shanghai Astronomical Observatory, United Kingdom Participation Group, Universidad Nacional Autónoma de México, University of Arizona, University of Colorado Boulder, University of Oxford, University of Portsmouth, University of Utah, University of Virginia, University of Washington, University of Wisconsin, Vanderbilt University, and Yale University.


The Legacy Surveys consist of three individual and complementary projects: the Dark Energy Camera Legacy Survey (DECaLS; Proposal ID \#2014B-0404; PIs: David Schlegel and Arjun Dey), the Beijing-Arizona Sky Survey (BASS; NOAO Prop. ID \#2015A-0801; PIs: Zhou Xu and Xiaohui Fan), and the Mayall z-band Legacy Survey (MzLS; Prop. ID \#2016A-0453; PI: Arjun Dey). DECaLS, BASS and MzLS together include data obtained, respectively, at the Blanco telescope, Cerro Tololo Inter-American Observatory, NSF’s NOIRLab; the Bok telescope, Steward Observatory, University of Arizona; and the Mayall telescope, Kitt Peak National Observatory, NOIRLab. Pipeline processing and analyses of the data were supported by NOIRLab and the Lawrence Berkeley National Laboratory (LBNL). The Legacy Surveys project is honored to be permitted to conduct astronomical research on Iolkam Du’ag (Kitt Peak), a mountain with particular significance to the Tohono O’odham Nation.

NOIRLab is operated by the Association of Universities for Research in Astronomy (AURA) under a cooperative agreement with the National Science Foundation. LBNL is managed by the Regents of the University of California under contract to the U.S. Department of Energy.

This project used data obtained with the Dark Energy Camera (DECam), which was constructed by the Dark Energy Survey (DES) collaboration. Funding for the DES Projects has been provided by the U.S. Department of Energy, the U.S. National Science Foundation, the Ministry of Science and Education of Spain, the Science and Technology Facilities Council of the United Kingdom, the Higher Education Funding Council for England, the National Center for Supercomputing Applications at the University of Illinois at Urbana-Champaign, the Kavli Institute of Cosmological Physics at the University of Chicago, Center for Cosmology and Astro-Particle Physics at the Ohio State University, the Mitchell Institute for Fundamental Physics and Astronomy at Texas A\&M University, Financiadora de Estudos e Projetos, Fundacao Carlos Chagas Filho de Amparo, Financiadora de Estudos e Projetos, Fundacao Carlos Chagas Filho de Amparo a Pesquisa do Estado do Rio de Janeiro, Conselho Nacional de Desenvolvimento Cientifico e Tecnologico and the Ministerio da Ciencia, Tecnologia e Inovacao, the Deutsche Forschungsgemeinschaft and the Collaborating Institutions in the Dark Energy Survey. The Collaborating Institutions are Argonne National Laboratory, the University of California at Santa Cruz, the University of Cambridge, Centro de Investigaciones Energeticas, Medioambientales y Tecnologicas-Madrid, the University of Chicago, University College London, the DES-Brazil Consortium, the University of Edinburgh, the Eidgenossische Technische Hochschule (ETH) Zurich, Fermi National Accelerator Laboratory, the University of Illinois at Urbana-Champaign, the Institut de Ciencies de l’Espai (IEEC/CSIC), the Institut de Fisica d’Altes Energies, Lawrence Berkeley National Laboratory, the Ludwig Maximilians Universitat Munchen and the associated Excellence Cluster Universe, the University of Michigan, NSF’s NOIRLab, the University of Nottingham, the Ohio State University, the University of Pennsylvania, the University of Portsmouth, SLAC National Accelerator Laboratory, Stanford University, the University of Sussex, and Texas A\&M University.

BASS is a key project of the Telescope Access Program (TAP), which has been funded by the National Astronomical Observatories of China, the Chinese Academy of Sciences (the Strategic Priority Research Program “The Emergence of Cosmological Structures” Grant \#XDB09000000), and the Special Fund for Astronomy from the Ministry of Finance. The BASS is also supported by the External Cooperation Program of Chinese Academy of Sciences (Grant \#114A11KYSB20160057), and Chinese National Natural Science Foundation (Grant \#12120101003, \#11433005).

The Legacy Survey team makes use of data products from the Near-Earth Object Wide-field Infrared Survey Explorer (NEOWISE), which is a project of the Jet Propulsion Laboratory/California Institute of Technology. NEOWISE is funded by the National Aeronautics and Space Administration.

The Legacy Surveys imaging of the DESI footprint is supported by the Director, Office of Science, Office of High Energy Physics of the U.S. Department of Energy under Contract No. DE-AC02-05CH1123, by the National Energy Research Scientific Computing Center, a DOE Office of Science User Facility under the same contract; and by the U.S. National Science Foundation, Division of Astronomical Sciences under Contract No. AST-0950945 to NOAO.


This project used public archival data from the Dark Energy Survey (DES). Funding for the DES Projects has been provided by the U.S. Department of Energy, the U.S. National Science Foundation, the Ministry of Science and Education of Spain, the Science and Technology Facilities Council of the United Kingdom, the Higher Education Funding Council for England, the National Center for Supercomputing Applications at the University of Illinois at Urbana–Champaign, the Kavli Institute of Cosmological Physics at the University of Chicago, the Center for Cosmology and Astro-Particle Physics at the Ohio State University, the Mitchell Institute for Fundamental Physics and Astronomy at Texas A\&M University, Financiadora de Estudos e Projetos, Fundação Carlos Chagas Filho de Amparo à Pesquisa do Estado do Rio de Janeiro, Conselho Nacional de Desenvolvimento Científico e Tecnológico and the Ministério da Ciência, Tecnologia e Inovação, the Deutsche Forschungsgemeinschaft and the Collaborating Institutions in the Dark Energy Survey.

The Collaborating Institutions are Argonne National Laboratory, the University of California at Santa Cruz, the University of Cambridge, Centro de Investigaciones Enérgeticas, Medioambientales y Tecnológicas–Madrid, the University of Chicago, University College London, the DES-Brazil Consortium, the University of Edinburgh, the Eidgenössische Technische Hochschule (ETH) Zürich, Fermi National Accelerator Laboratory, the University of Illinois at Urbana-Champaign, the Institut de Ciències de l’Espai (IEEC/CSIC), the Institut de Física d’Altes Energies, Lawrence Berkeley National Laboratory, the Ludwig-Maximilians Universität München and the associated Excellence Cluster Universe, the University of Michigan, the National Optical Astronomy Observatory, the University of Nottingham, The Ohio State University, the OzDES Membership Consortium, the University of Pennsylvania, the University of Portsmouth, SLAC National Accelerator Laboratory, Stanford University, the University of Sussex, and Texas A\&M University.

Based in part on observations at Cerro Tololo Inter-American Observatory, National Optical Astronomy Observatory, which is operated by the Association of Universities for Research in Astronomy (AURA) under a cooperative agreement with the National Science Foundation.

\section*{Data Availability}

The data produced for this article are available in its online edition.

The WiggleZ catalogue is available at \url{https://doi.org/10.1093/mnras/stx2963}, and its spectra at \url{https://rdm.uq.edu.au/files/17729f20-50a0-11ec-97fe-7bcb2225c4dd}.

ASKAP-FLASH Pilot Survey data are available at \url{https://data.csiro.au/collection/csiro:47965}.



\bibliographystyle{mnras}
\bibliography{paper_bib} 




\appendix

\section{Data}\label{data}

This work produced data via cross-matching of WiggleZ data, and additional data assembled to match the WiggleZ survey's construction, with FLASH radio data. An associated and an offset dataset are produced, by cross-matching within two different radii; associated data with a radius of 5\,arcseconds or less, and offset data with a radius falling between 5\,arcseconds and 20. After this process, the data is then run through an automated pipeline built for this work, which then attempts to isolate the H$\upalpha$, H$\upbeta$, {[}OIII{]}, and {[}NII{]} spectral lines from an object's spectrum, and extract a value for EW for all lines.

Table \ref{tab:finalised_associated_data} consists of many columns and entries, and is a sample of 5 objects found in the associated dataset, demonstrating how different entries can look depending on their field, the spectroscopic survey they are from, and whether spectral line data was possible to obtain. The write-up on the associated table below is relevant to the offset data also, as the kind of contents in each column is identical.

Further information on WiggleZ data can be found in \citet{Drinkwater2018}. For ASKAP survey data in general, refer to \citet{Hotan2021}.


\subsection{Cross-Matched Data}\label{cross_matched_cols}

This is the data that is fed into the pipeline to obtain spectroscopic data from the FITS files. It contains a restricted sample where H$\upalpha$, H$\upbeta$, {[}OIII{]}, or {[}NII{]} lines are detected. A spectrum does not need all four lines to be present for its data to be extracted, however depending on which lines are not present, it may result in an object not being plotted in one or both of the BPT and WHAN diagrams.

\medskip
\noindent
\textbf{Column 1: Object name.} Name of the WiggleZ-type object as appears in the catalogue that the object is sourced from. If no name was available in the source catalogue, then name was generated.

\smallskip
\noindent
\textbf{Column 2 and 3: Right ascension and declination of WiggleZ-type object.} The RA and Dec of the object in degrees in J2000 coordinates.

\smallskip
\noindent
\textbf{Column 4: Redshift of object.} The spectroscopic redshift of the WiggleZ-type object was obtained from source catalogue (see column 23: 'z survey').

\smallskip
\noindent
\textbf{Column 5: WiggleZ Redshift quality.} The quality of redshift assigned to each WiggleZ object. This is obtained when the redshift that \verb\RUNZ\ assigns is manually checked. The possible QOP values that can be assigned range from 1 to 5, with 5 being very high certainty redshift, and 1 indicating that it was not possible to assign a redshift. As other spectroscopic redshifts assign QOP values differently, those have been retained in column 26 for BOSS/eBOSS.

\smallskip
\noindent
\textbf{Column 6, 7, 8, 9, 10: FUV, NUV, \textit{g}, \textit{r}, and \textit{i} band photometry (mag).} Photometry for W09 and W15 objects is from the WiggleZ survey, while optical photometry for data in the extended sample is taken from DES, and NUV and FUV photometry from GALEX.

\smallskip
\noindent
\textbf{Column 11: WiggleZ object class} Class of WiggleZ object.

\smallskip
\noindent
\textbf{Column 12: Absolute FUV magnitude.} Only WiggleZ objects possess this data. For further detail, see \citet{Drinkwater2018}.

\smallskip
\noindent
\textbf{Column 13: Stellar mass.} Calculated stellar mass of galaxy, from WiggleZ survey.

\smallskip
\noindent
\textbf{Column 14: SpecFile.} Name of specific FITS spectrum file associated with each object, taken from WiggleZ, BOSS, eBoss-ELG, and eBOSS-LRG surveys.

\smallskip
\noindent
\textbf{Column 15: Comment.} Any comment made on the object in question, from WiggleZ or BOSS/eBOSS. This can be about object type, or spectral feature.

\smallskip
\noindent
\textbf{Column 16: FLASH island ID.} ID of radio object from FLASH survey. Contains 7 digit ASKAP SBID and 4 digit ASKAP catalogue ID.

\smallskip
\noindent
\textbf{Column 17: FLASH island name.} Name of radio object from FLASH survey. This is produced automatically during the ASKAP data processing pipeline from the sexagesimal RA and Dec.

\smallskip
\noindent
\textbf{Column 18: Number of components.} Number of radio components the \verb\Selavy\ algorithm associates together in 3D space to produce the FLASH island.

\smallskip
\noindent
\textbf{Column 19 and 20: Right ascension and declination (of radio source).} The J2000 coordinates RA and Dec of the cross-matched radio source in decimal degrees.

\smallskip
\noindent
\textbf{Column 21: Integrated flux density (mJy).} The integrated flux density of the cross-matched FLASH radio source.

\smallskip
\noindent
\textbf{Column 22: Peak flux density (mJy\,beam$^{-1}$).} The peak flux density of the cross-matched FLASH radio source.

\smallskip
\noindent
\textbf{Column 24: Field.} The survey area that the optical object is in. `W09' and `W15' refer to the WiggleZ data that lie within the WiggleZ 9h and 15h fields. `Ext' refers to objects in the synthesised WiggleZ-type catalogue that lie in an equatorial strip between 21h and 3h.

\smallskip
\noindent
\textbf{Column 25: Field.} Separation in arcseconds between the optical object and the FLASH radio object it has been cross-matched with. All associated objects will have values of 5 or less. Offset objects will have between 5 and 20.

\smallskip
\noindent
\textbf{Column 26: DESI DR9 Legacy Survey ID.} The ID of the optical object taken from the 9th data release of the DESI spectroscopic redshift Legacy Survey. Only extended sample objects possess this.

\smallskip
\noindent
\textbf{Column 27: zwarning.} Quality flag for BOSS/eBOSS spectra. 0 corresponds to a good quality spectrum that a redshift can be obtained for with high confidence. Other flags can be found on \citet{SDSS}, however all BOSS/eBOSS objects in this work have the 0 flag.

\smallskip
\noindent
\textbf{Column 28 and 29: Right ascension and declination (of FIRST object radio source).} The J2000 coordinates RA and Dec of the cross-matched FIRST radio source in decimal degrees. This is done with the intention of obtaining more precise radio positions on the sky. However, not every object in this catalogue that has been cross-matched against a FLASH radio source will have been observed by the FIRST survey.

\smallskip
\noindent
\textbf{Column 30, 31: H$\upalpha$, equivalent width and associated error (\AA).} These are obtained for this work by passing spectra through an automated pipeline, and are only produced if the SNR exceeds 3. Not all objects will possess all or even some of the data, due to redshift, spectra quality, or lines simply not existing in the spectra. Error of EW values obtained via this work's automated pipeline.

\smallskip
\noindent
\textbf{Column 32, 33, 34, and 35: H$\upalpha$, H$\upbeta$, {[}OIII{]}, and {[}NII{]} line fluxes ($\mathrm{e^{-16}\,erg\,}$$\AA\,$$\mathrm{cm^{-2}\,s^{-1}}$)}. Integrated fluxes of H$\upalpha$, H$\upbeta$, {[}OIII{]}, and {[}NII{]}. These are obtained for this work by passing spectra through an automated pipeline, and are only produced if the SNR exceeds 3. Not all objects will possess all or even some of the data, due to redshift, spectra quality, or lines simply not existing in the spectra.

\smallskip
\noindent
\textbf{Column 36, 37, 38, 39: H$\upalpha$, H$\upbeta$, {[}OIII{]}, and {[}NII{]} line flux errors (\AA).} Error of line flux values obtained via this work's automated pipeline.


\subsection{Spectral Data}\label{spectral_fits_files}

The specific FITS file designation for each object is found in the SpecFile column of both the associated and offset tables (see Table \ref{tab:finalised_associated_data}).

\begin{figure*}
    \includegraphics[width=\linewidth]{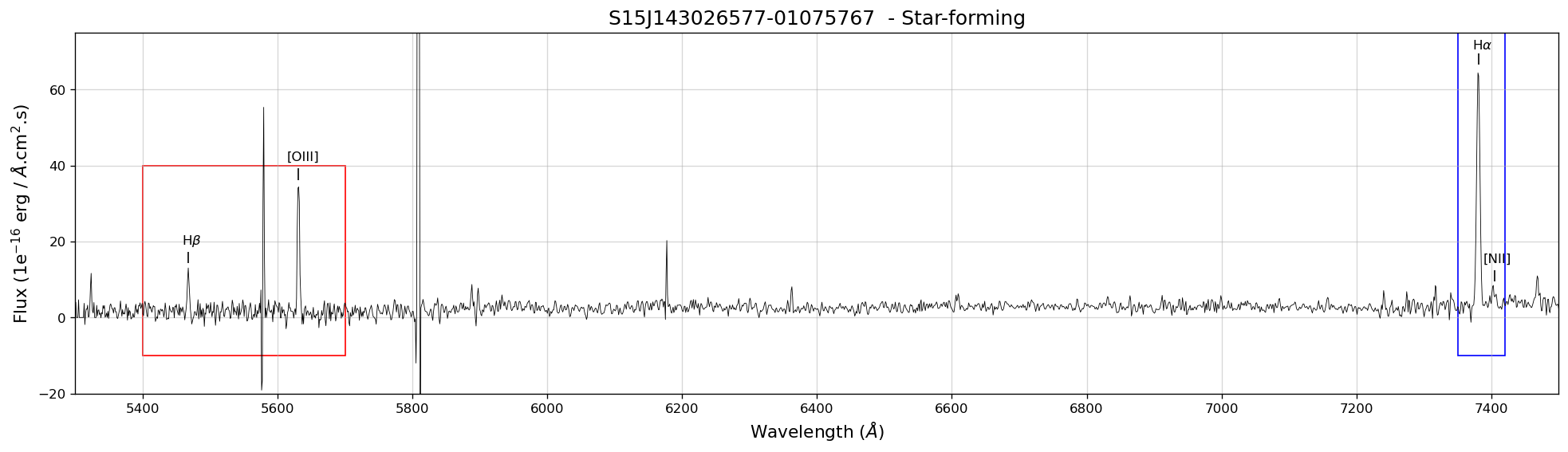}
    \par 
\begin{multicols}{2}
    \includegraphics[width=\linewidth]{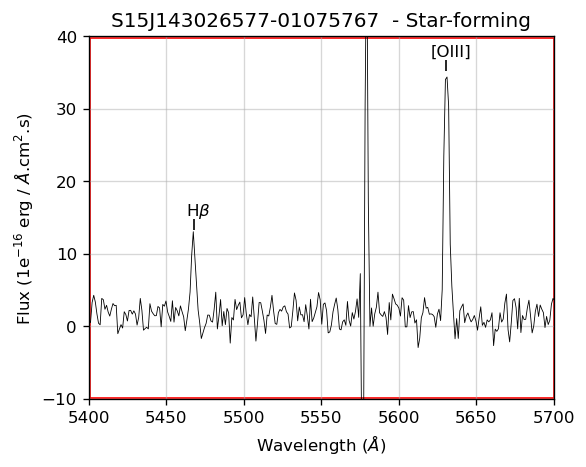}
    \par
    \includegraphics[width=\linewidth]{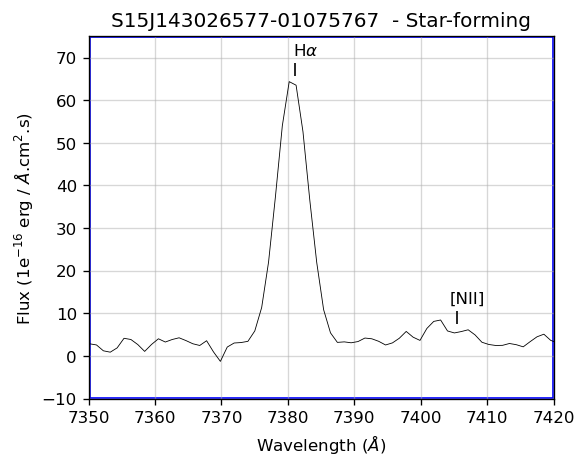}
    \par
    \end{multicols}
\caption{Top: Example spectrum of star-forming WiggleZ galaxy S15J143453974+00215187 as classified using BPT, after continuum normalisation. Its redshift is z = 0.12. Cutout areas are indicated with coloured boxes matching the relevant cutout. Bottom left: Cutout of star-forming spectrum showing H$\upbeta$ and [OIII] lines. Bottom right: Cutout showing H$\upalpha$ and [NII] lines. Note the different flux and wavelength scales for the two cutouts. Examples of `bad pixels' are displayed at $\sim{5575}$\,\AA, and $\sim{5800}$\,\AA. They are separate enough from the peaks that they do not pose an issue for this spectrum.}
\label{fig:sf_spectrum}
\end{figure*}

\begin{figure*}
    \includegraphics[width=\linewidth]{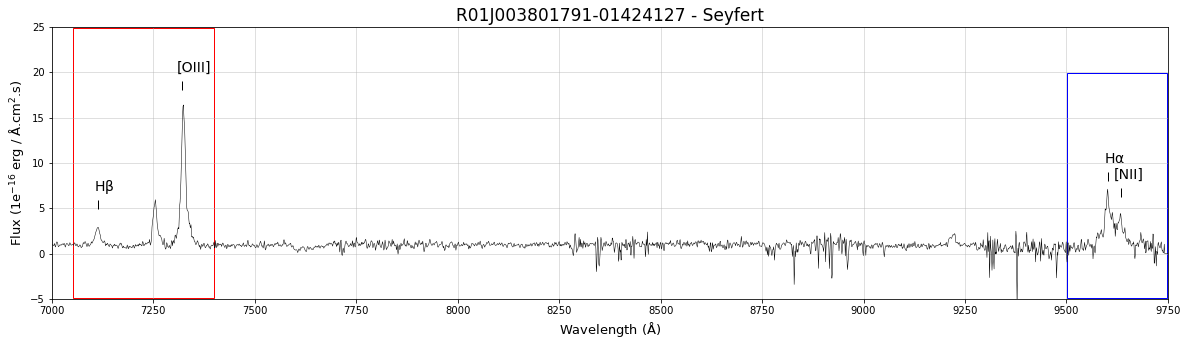}
    \par 
\begin{multicols}{2}
    \includegraphics[width=\linewidth]{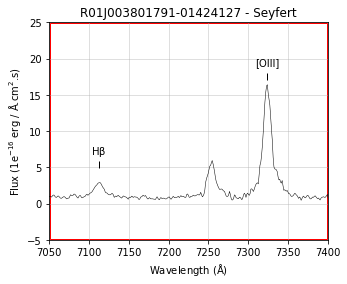}
    \par
    \includegraphics[width=\linewidth]{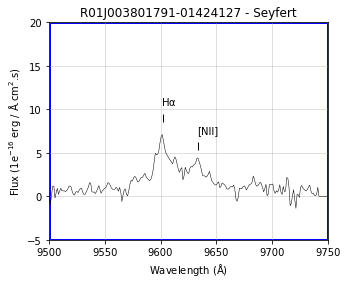}
    \par
    \end{multicols}
\caption{Top: Example spectrum of Seyfert WiggleZ galaxy R01J003801791-01424127 as classified using BPT, after continuum normalisation, with a redshift of z = 0.46. Cutout areas are indicated with coloured boxes matching the relevant cutout. Bottom left: Cutout of Seyfert spectrum showing H$\upbeta$ and [OIII] lines. Bottom right: Cutout showing H$\upalpha$ and [NII] lines.}
\label{fig:seyfert_spectrum}
\end{figure*}

\begin{figure*}
    \includegraphics[width=\linewidth]{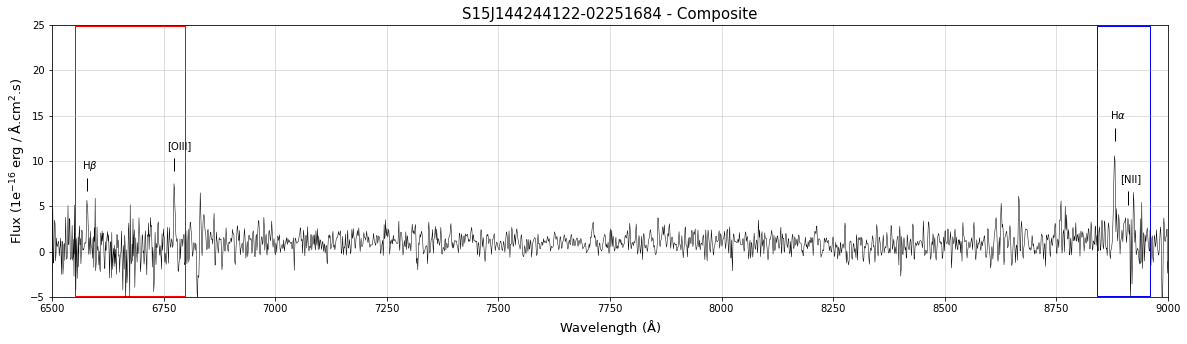}
    \par 
\begin{multicols}{2}
    \includegraphics[width=\linewidth]{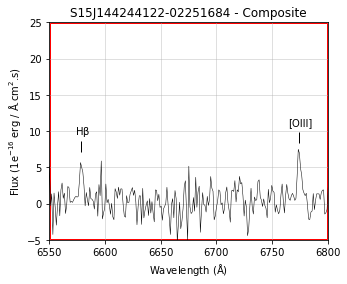}
    \par
    \includegraphics[width=\linewidth]{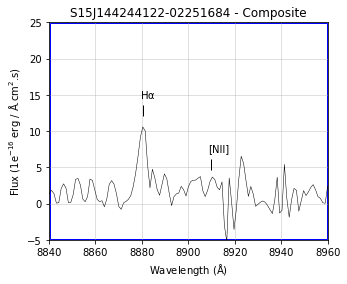}
    \par
    \end{multicols}
\caption{Top: Example spectrum of composite WiggleZ galaxy S15J144244122-02251684 as classified using BPT, after continuum normalisation, with a redshift of z = 0.35. Cutout areas are indicated with coloured boxes matching the relevant cutout. Bottom left: Cutout of composite spectrum showing H$\upbeta$ and [OIII] lines. Bottom right: Cutout showing H$\upalpha$ and [NII] lines. It is important to note that the level of noise is not higher in this spectrum, but the line intensities are smaller.}
\label{fig:comp_spectrum}
\end{figure*}

Example spectra for objects found in this paper are included as Figures \ref{fig:sf_spectrum}, \ref{fig:seyfert_spectrum}, and \ref{fig:comp_spectrum}. Note that some spectra in the source surveys are subject to `bad pixels' and the effects of cosmic rays. These should not have an effect on the data passed through the pipeline for spectral lines, as only specific parts of the spectra are of interest, and mechanisms are in place to deal with anomalous results from the spectral lines. Standard cutoffs such as rejection of lines that do not exceed SNR of 3 are used. Detections assigned EWs with a SNR greater than 3, but that would imply detections of peaks that are very broad but with peak amplitudes less than 0.1\,$\mathrm{e^{-16}\,erg\,}$\AA\,$\mathrm{cm^{-2}\,s^{-1}}$, i.e. comparative to noise in the spectrum, or with very large changes in amplitude but with widths of less than 5 pixels, i.e. bad pixels/cosmic rays, are rejected.


\FloatBarrier

\begin{table*}
    \centering
    \caption{Sample of five rows of associated (<\,5\,arcsecond separation between optical object and FLASH object) cross-matched data, arranged by ascending RA. Column order is identical for offset objects. Full tables are available online.}\label{tab:finalised_associated_data}
    \begin{filecontents*}{sample_associated.csv}
Name,RAJ2000,DEJ2000,z,q_z,FUV,NUV,gmag,rmag,imag,Class,MFUV,Mass,SpecFile,Com,island_id,island_name,n_components,ra_deg_cont,dec_deg_cont,flux_int,flux_peak,z_survey,field,separation(arcsec),ls_id,zwarning,RA_FIRST,DEC_FIRST,Ha_EW,e_Ha_EW,Ha_LF,HB_LF,OIII_LF,NII_LF,e_Ha_LF,e_HB_LF,e_OIII_LF,e_NII_LF
R01J005151101-01402649,12.962941,-1.67408,0.131,4,,22.34,20.8,20.25,19.97,WIG_RCS2,-16.78, ,wig011695.fits, ,SB42300_island_5406,J005151-014030,1,12.963026,-1.675037,0.931,1.319,WiggleZ,Ext,3.3136,'9906620760004906',,,,-28,1.28,587.3,122.8,249.9,109.5,26.37,23.64,25.58,7.63
R01J005358363-01333111,13.493243,-1.558635,0.404,4,23.58,22.52,21.66,20.83,20.69,WIG_RCS2,-19.13,9.54,wig012834.fits, ,SB42300_island_5528,J005358-013335,1,13.492535,-1.559955,0.962,1.279,WiggleZ,Ext,4.7386,'9906620854509938',,,,-87.3,4.36,1098.1,207.7,273.4,259.2,47.39,10.51,7.69,10.17
S09J085802335+02462617,134.509729,2.773936,0.246,3,25.43,22.06,20.72,20.89,20.38,WIG_SDSS,-18.08, ,wig047052.fits, ,SB34559_island_10767,J085802+024625,1,134.509299,2.773653,0.406,0.56,WiggleZ,W09,2.3352,,,,,-4.8,0.22,104.2,19.2,35.5,-45.7,4.03,11.8,12.76,5.81
S15J142503516+01372061,216.264651,1.622392,0.446,4,23.64,21.51,23.09,21.39,20.94,WIG_SDSS,-20.03,10.61,wig130560.fits, ,SB41085_island_1906,J142503+013720,1,216.263762,1.622482,4.714,4.855,WiggleZ,W15,3.2983,,,216.263887,1.622367,-34,1.17,412.4,94.3,327.4,114.1,28.64,3.46,6.71,9
S15J142555504-00512709,216.481266,-0.857524,1.025,3,,22.5,22.12,22.02,21.71,WIG_SDSS,-21.07,8.99,wig131355.fits, ,SB41085_island_382,J142555-005126,3,216.481823,-0.857372,31.878,30.013,WiggleZ,W15,1.8053,,,216.481839,-0.857261,,,,,,,,,,
\end{filecontents*}
\csvreader[tabular=lrrrrrrrrrrr,
    table head=\hline \hline Name & RAJ2000 & DEJ2000 & z & QOP & FUV & NUV & g & r & i & class & MFUV\\
    1 & 2 & 3 & 4 & 5 & 6 & 7 & 8 & 9 & 10 & 11 & 12
    \\\hline,
    filter={\value{csvrow}<5},
    respect all=true,
    table foot=\hline,]
    {sample_associated.csv}{}%
    {\csvcoli & \csvcolii & \csvcoliii & \csvcoliv & \csvcolv & \csvcolvi & \csvcolvii & \csvcolviii & \csvcolix & \csvcolx & \csvcolxi & \csvcolxii}%
\end{table*}

\addtocounter{table}{-1}
\begin{table*}
    \centering
    \csvreader[tabular=rrrrrrrrr,
    table head=\hline \hline mass & SpecFile & comment & FLASH island ID & island name & no. components & FLASH RA & FLASH Dec\\
    13 & 14 & 15 & 16 & 17 & 18 & 19 & 20
    \\\hline,
    filter={\value{csvrow}<5},
    respect all=true,
    table foot=\hline,]
    {sample_associated.csv}{}%
    {\csvcolxiii & \csvcolxiv & \csvcolxv & \csvcolxvi & \csvcolxvii & \csvcolxviii & \csvcolxix & \csvcolxx}%
\end{table*}

\addtocounter{table}{-1}
\begin{table*}
    \centering
    \csvreader[tabular=rrrrrrrrr,
    table head=\hline \hline int. flux & peak flux & z survey & field & separation & DESI LS ID & zwarning & FIRST RA & FIRST Dec \\
    21 & 22 & 23 & 24 & 25 & 26 & 27 & 28 & 29
    \\\hline,
    filter={\value{csvrow}<5},
    respect all=true,
    table foot=\hline,]
    {sample_associated.csv}{}%
    {\csvcolxxi & \csvcolxxii & \csvcolxxiii & \csvcolxxiv & \csvcolxxv & \csvcolxxvi & \csvcolxxvii & \csvcolxxviii & \csvcolxxix}%
\end{table*}

\addtocounter{table}{-1}
\begin{table*}\label{}
    \centering
    \csvreader[tabular=rrrrrrrrrrrrr,
    table head=\hline \hline H$\mathrm{\upalpha}$ EW & $\mathrm{\upDelta H\upalpha}$ EW & $\mathrm{H\upalpha}$ LF & $\mathrm{H\upbeta}$ LF & $\mathrm{[OIII]}$ LF & $\mathrm{[NII]}$ LF & $\mathrm{\upDelta H\upalpha}$ LF & $\mathrm{\upDelta H\upbeta}$ LF & $\mathrm{\upDelta [OIII]}$ LF & $\mathrm{\upDelta [NII]}$ LF\\
    30 & 31 & 32 & 33 & 34 & 35 & 36 & 37 & 38
    \\\hline,
    filter={\value{csvrow}<5},
    respect all=true,
    table foot=\hline,]
    {sample_associated.csv}{}%
    {\csvcolxxx & \csvcolxxxi & \csvcolxxxii & \csvcolxxxiii & \csvcolxxxiv & \csvcolxxxv & \csvcolxxxvi & \csvcolxxxvii & \csvcolxxxviii & \csvcolxxxix}%
\end{table*}


\bsp	
\label{lastpage}
\end{document}